\documentclass[11pt,a4paper]{article}
\usepackage[utf8]{inputenc}
\usepackage[T1]{fontenc}
\usepackage[english]{babel}
\usepackage[a4paper,top=2.5cm,bottom=2.8cm,left=2.7cm,right=2.7cm]{geometry}
\usepackage{setspace}
\usepackage{mathpazo}      % Palatino-like text + math (matches Apollo paper)
\usepackage{microtype}
\usepackage{graphicx}
\usepackage{float}
\usepackage{xcolor}
\definecolor{apolloblue}{HTML}{1F3A8A}
\usepackage{caption}
\usepackage{enumitem}
\setlist{itemsep=2pt,topsep=4pt,parsep=0pt}
\usepackage[colorlinks=true,
            linkcolor=black,
            urlcolor=blue!55!black,
            citecolor=blue!55!black,
            filecolor=blue!55!black,
            pdfauthor={Matteo Pistillo},
            pdftitle={Internal Deployment in the AI Act}]{hyperref}
\usepackage{url}
\usepackage[hang,flushmargin,bottom]{footmisc}

\usepackage{titlesec}
\titleformat{\section}{\normalfont\Large\bfseries}{}{0pt}{}
\titleformat{\subsection}{\normalfont\large\bfseries}{}{0pt}{}
\titleformat{\subsubsection}{\normalfont\normalsize\bfseries}{}{0pt}{}
\titlespacing*{\section}{0pt}{2.4ex plus 1ex minus .2ex}{1.6ex plus .2ex}
\titlespacing*{\subsection}{0pt}{2.0ex plus 1ex minus .2ex}{1.2ex plus .2ex}
\titlespacing*{\subsubsection}{0pt}{1.6ex plus .8ex minus .2ex}{0.9ex plus .2ex}
\usepackage{fancyhdr}
\fancypagestyle{titlepg}{%
  \fancyhf{}%
}
\begin{document}
\thispagestyle{titlepg}
% --- Title block (Apollo-style) --------------------------------------------
\begin{center}
\vspace*{1.5em}
{\fontsize{17pt}{21pt}\selectfont\bfseries Internal Deployment in the AI Act\par}
\vspace{2.0em}
{\large\bfseries Matteo Pistillo\textsuperscript{*}\textsuperscript{\dag}\par}
\vspace{0.6em}
{\normalsize Apollo Research\par}
\vspace{0.4em}
\end{center}
% --- Title footnotes (symbol markers, do not consume numbered slots) -------
\begingroup
\renewcommand{\thefootnote}{\fnsymbol{footnote}}
\footnotetext[1]{Pre-print. Written for the \emph{Cambridge Commentary on EU General-Purpose AI Law}.}
\footnotetext[2]{Correspondence to \href{mailto:matteo@apolloresearch.ai}{matteo@apolloresearch.ai}.}
\endgroup
\setcounter{footnote}{0}
\vspace{1.5em}

% your content starts here...

\begin{abstract}
\noindent
This memorandum analyzes and stress-tests arguments in favor and against the inclusion of \emph{internal deployment} within the scope of the European Union (EU) Artificial Intelligence (AI) Act. Specifically, this memorandum first analyzes interpretative pathways based on Article 2(1)(a)--(c) supporting the application of the AI Act to internally deployed AI models and systems (\S2). Then, it examines possible objections and exceptions based on Articles 2(6) and 2(8), with particular attention to the complexity of the scientific R\&D exception under Article 2(6) (\S3). Finally, it illustrates how Articles 2(1), 2(6), and 2(8) can be viewed as complementary to each other, once broken down to their most plausible meaning and interpreted in conjunction with Articles 3(1), 3(3), 3(4), 3(9), 3(10), 3(11), 3(12), 3(63), and Recitals 12, 13, 21, 25, 97, 109, and 110 (\S4; Figure~\ref{fig:scope1}).

\begin{figure}[H]
\centering
\includegraphics[width=0.65\textwidth]{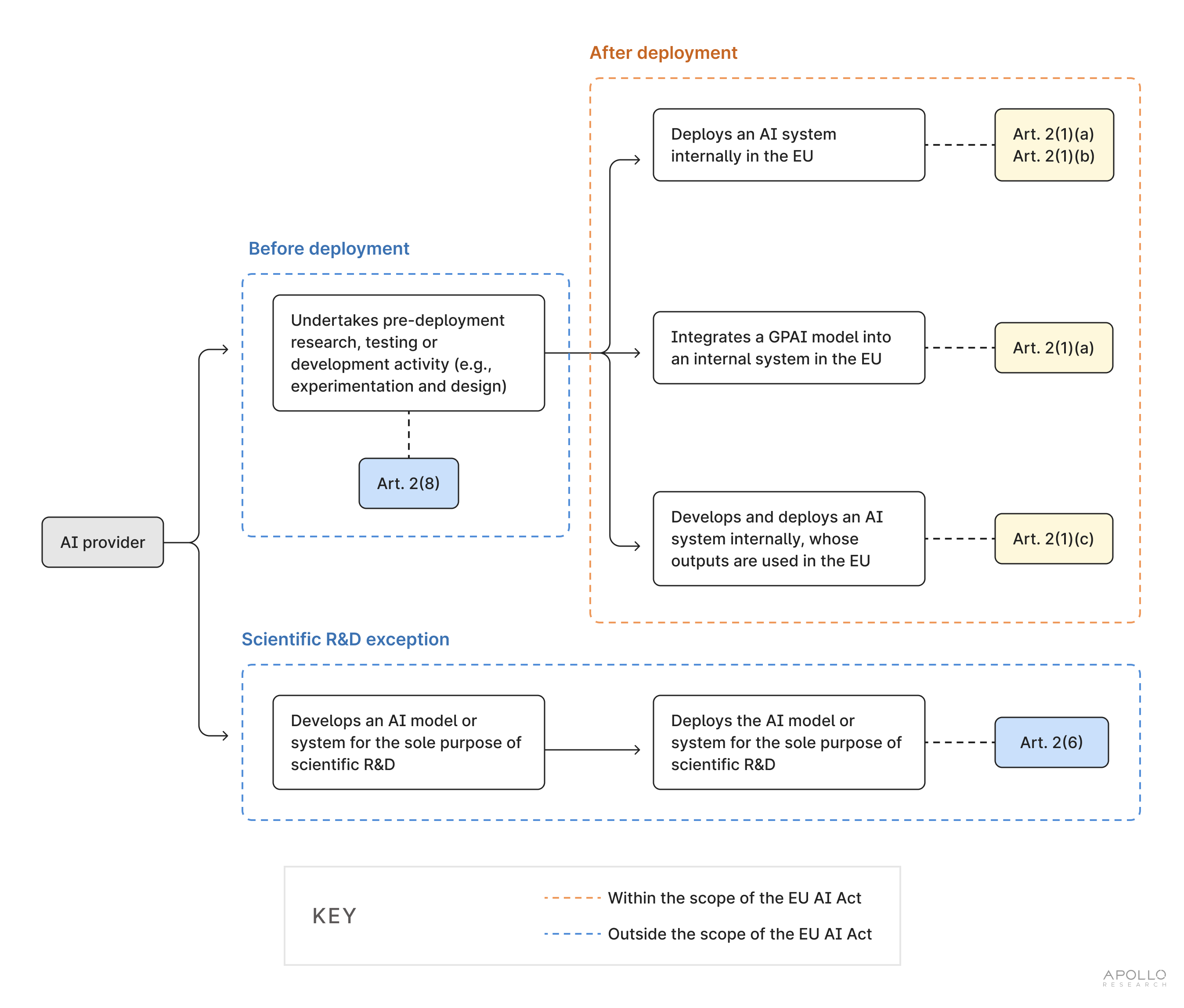}
\caption{Visualization of internal AI models and systems arguably falling \emph{within} (in yellow) and \emph{outside} (in blue) the scope of the AI Act. In summary, internal deployment triggers the application of the AI Act (Article 2(1)(a)--(c)) unless an internal AI model or system is specifically developed and deployed for the sole purpose of scientific R\&D (Article 2(6)). Pre-internal-deployment research, testing and development activity arguably falls outside the scope of the AI Act (Article 2(8)).}
\label{fig:scope1}
\end{figure}
\end{abstract} 

\subsection*{1. Introduction}
\addcontentsline{toc}{section}{1. Introduction}

This memorandum focuses on one of the most challenging questions surrounding the European Union (EU) Artificial Intelligence (AI) Act:\footnote{Regulation (EU) 2024/1689 of June 13, 2024 (AI Act) (\href{https://eur-lex.europa.eu/legal-content/EN/TXT/PDF/?uri=OJ:L_202401689}{AI Act}).} whether and to what extent the AI Act applies to what frontier AI companies and researchers refer to as ``internal deployment.''\footnote{\emph{See} \href{https://arxiv.org/abs/2504.12170}{Stix et al.}, 2025; \href{https://static1.squarespace.com/static/64edf8e7f2b10d716b5ba0e1/t/687e324254b8df665abc5664/1753100867033/Managing+Risks+from+Internal+AI+Systems.pdf}{Acharya and Delaney}, 2025; \href{https://stevenadler.substack.com/p/ai-companies-unmonitored-internal}{Adler}, 2025; \href{https://internationalaisafetyreport.org/sites/default/files/2025-10/international_ai_safety_report_2025_english.pdf}{International Scientific Report on the Safety of Advanced AI}, 2025, at 35 (``Deployment can take several forms: internal deployment for use by the system's developer, or external deployment either publicly or to private customers''); \href{https://www-cdn.anthropic.com/08ab9158070959f88f296514c21b7facce6f52bc.pdf}{Claude Mythos Preview System Card}, 2026, at 57, 61, 62, 65, 132, 203; \href{https://deploymentsafety.openai.com/gpt-5-5/gpt-5-5.pdf}{GPT-5.5 System Card}, 2026, at 13, 14, 43; \href{https://ai.meta.com/static-resource/Meta_Advanced-AI-Scaling-Framework-v2}{Meta Advanced AI Scaling Framework}, 2026, at 4, 9, 18, 27, 42.} The inclusion of internal deployment within the scope of the AI Act remains an open question.\footnote{\emph{See} \href{https://arxiv.org/abs/2504.12170}{Stix et al.}, 2025, at 27.} This memorandum attempts to unpack this open question, offering an overview of several potential interpretative pathways. Specifically, \S{}1 of this memorandum  provides an initial introduction to the concept of internal deployment. Next, \S{}2 grounds itself in the technical context of frontier AI development and deployment and elaborates on potential arguments supporting the application of the AI Act to internal deployment. Finally, \S{}3 outlines arguments for opposing, or at least restricting, the application of the AI Act to internal deployment, as well as potential counterarguments.

Internal deployment is an umbrella term used to describe situations in which \textbf{an AI provider develops an AI model or system and makes it available for use within the provider itself}.\footnote{This definition is adapted from \href{https://arxiv.org/abs/2504.12170}{Stix et al.}, 2025, at 6 (defining ``internal deployment'' as ``the act of making an AI system available for access and/or usage exclusively for the developing organization''). \emph{See also} \href{https://static1.squarespace.com/static/64edf8e7f2b10d716b5ba0e1/t/687e324254b8df665abc5664/1753100867033/Managing+Risks+from+Internal+AI+Systems.pdf}{Acharya and Delaney}, 2025 (using the term ``internal deployment,'' and defining ``internal models'' as ``AI systems that are only accessible to company employees (and perhaps to a few external experts given access by the company)''); \href{https://ai.meta.com/static-resource/Meta_Advanced-AI-Scaling-Framework-v2}{Meta Advanced AI Scaling Framework}, 2026, at 42 (defining ``internal deployment'' as ``models that are exclusively available to Meta personnel''). Some researchers use the term ``internal usage'' or ``internal use'' to refer to ``internal deployment.'' \emph{See}, for instance, \href{https://metr.org/blog/2025-01-17-ai-models-dangerous-before-public-deployment/}{METR}, 2025 (referring to internal deployment through the expression ``internal usage''); \href{https://stevenadler.substack.com/p/ai-companies-unmonitored-internal}{Adler}, 2025 (defining ``a company's internal use of AI'' as ``internal deployment''). In this respect, it is important to observe that the meaning of ``internal deployment,'' as commonly used by AI companies and researchers, is generally broader than the definition of ``deployer'' contained in Article 3(4) and Recital 13 of the AI Act.} In other words, internal deployment describes the behavior of an AI provider that, instead of publicly releasing an AI model or systems, elects to use it only within the provider walls. As an umbrella term, internal deployment covers a range of different scenarios diverging, among other aspects, in the exclusivity of access to AI models or systems, in the usage of these AI models or systems, and in the level of AI model or system autonomy.\footnote{\emph{See} \href{https://arxiv.org/abs/2504.12170}{Stix et al.}, 2025, at 12-14 (outlining various potential internal uses as well as various potential internal user groups).} An AI provider's exclusive access to and/or usage of an internal AI model or AI system can be temporary, or can also be permanent.\footnote{\emph{See} \href{https://arxiv.org/abs/2504.12170}{Stix et al.}, 2025, at 8.} In other words, internal AI models and systems may be first deployed internally, and then made available to the public;\footnote{For instance, OpenAI's model GPT-4 was available internally for six months before being publicly released. \emph{See} \href{https://cdn.openai.com/papers/gpt-4-system-card.pdf}{GPT-4 System Card}, 2023, at 59. Anthropic's Opus 4.6 and 4.7 and OpenAI's GPT-5.5 also appear to have been used internally before public release. \emph{See} \href{https://www-cdn.anthropic.com/6a5fa276ac68b9aeb0c8b6af5fa36326e0e166dd.pdf}{Claude Opus 4.6 System Card}, 2026, at 95 (``Throughout late-stage training, we deployed several snapshots of Claude Opus 4.6 for provisional internal use and evaluation, with increasingly broad uptake as time went on.''); \href{https://cdn.sanity.io/files/4zrzovbb/website/037f06850df7fbe871e206dad004c3db5fd50340.pdf}{Claude Opus 4.7 System Card}, 2026, at 95 (``We used versions of Claude Opus 4.7 substantially internally before deploying it''); \href{https://deploymentsafety.openai.com/gpt-5-5/gpt-5-5.pdf}{GPT-5.5 System Card}, 2026, at 14 (mentioning ``a pre-final version of GPT-5.5 internal usage'').} or they may be used exclusively within an AI provider for the entirety of their lifetime, and never made available to the public. Internal AI models and systems may be deployed to assist with, or automate, a variety of different tasks within a provider (for instance, within its HR, legal, or R\&D departments).\footnote{The usage of internal AI models and systems may span from light professional usage and work support (i.e., supporting a provider's staff with their workstream and outsourcing portions of existing or new workstreams, and then using the relevant outputs) to outright delegation to virtual co-workers or independent workers. \emph{See} \href{https://arxiv.org/abs/2504.12170}{Stix et al.}, 2025, at 12-14. \emph{See also} \href{https://stevenadler.substack.com/p/ai-companies-unmonitored-internal}{Adler}, 2025 (mentioning ``critical internal tasks like security reviews, interpretability analysis, and developing next-generation AI systems''); \href{https://arxiv.org/pdf/2305.15324}{Shevlane et al}., 2024, at 9; \href{https://static1.squarespace.com/static/64edf8e7f2b10d716b5ba0e1/t/687e324254b8df665abc5664/1753100867033/Managing+Risks+from+Internal+AI+Systems.pdf}{Acharya and Delaney}, 2025, 8-10.} The degree of autonomy of internal AI models and systems also sits on a spectrum, from undertakings that are small, self-contained projects, all the way to future arrangements that could engage ``a country of geniuses in a datacenter'' running very large segments of an AI provider's workflow autonomously.\footnote{\emph{See} \href{https://www.darioamodei.com/essay/machines-of-loving-grace}{Amodei}, 2024. Higher degrees of autonomy may entail less friction to speed, but also a lower degree of human oversight. \emph{See} \href{https://arxiv.org/abs/2504.12170}{Stix et al.}, at 16-17.}

One important, and potentially concerning,\footnote{In summary, concerns cluster around the risk that, once AI systems play a growing role in designing and training their successors, this could trigger a hyperbolic explosion in AI capabilities (\href{https://www.forethought.org/research/how-can-ai-labs-incorporate-risks-from-ai-accelerating-ai-progress-into}{Davidson}, 2025; \href{https://www.forethought.org/research/will-ai-r-and-d-automation-cause-a-software-intelligence-explosion}{Eth and Davidson}, 2025) and effectively prevent humans from understanding, auditing, or controlling the resulting technology (\href{https://ai-2027.com/}{Kokotajlo et al}., 2025). For an overview of these and other potential threats from internal deployment, \emph{see} \href{https://arxiv.org/abs/2504.12170}{Stix et al.}, 2025; \href{https://metr.org/blog/2025-01-17-ai-models-dangerous-before-public-deployment/}{METR}, 2025; and \href{https://static1.squarespace.com/static/64edf8e7f2b10d716b5ba0e1/t/687e324254b8df665abc5664/1753100867033/Managing+Risks+from+Internal+AI+Systems.pdf}{Acharya and Delaney}, 2025. \emph{See} \emph{also} a recent interview of Anthropic's co-founder and chief scientist, Jared Kaplan, who defined the decision to let AI systems train themselves as ``the biggest decision yet'' and ``the ultimate risk'' (\href{https://www.theguardian.com/technology/ng-interactive/2025/dec/02/jared-kaplan-artificial-intelligence-train-itself}{The Guardian}, 2025).} use that AI providers can make of their most capable AI models and systems is to use them to design, train, or improve future (and, plausibly, more capable) AI models and systems.\footnote{\emph{See} \href{https://alignment.anthropic.com/2026/auditing-overt-saboteur/}{Anthropic}, 2026, at 1 (``Anthropic uses existing Claude models to assist in the development of future generations of Claude models''); \href{https://www.forethought.org/research/three-types-of-intelligence-explosion}{Davidson et al.}, 2025; \href{https://www.forethought.org/research/will-ai-r-and-d-automation-cause-a-software-intelligence-explosion}{Eth and Davidson}, 2025. For clarity, AI R\&D is \emph{only one} of the possible internal uses of AI models and systems by AI providers. This clarification is important, as some provisions in the AI Act seem to concern internal deployment generally (\emph{see} \S{}\S{}2.1-3.1, and \S{}3.3 below), whereas other provisions seem to specifically concern \emph{only} AI R\&D (\emph{see} \S{}3.2 below). For the purposes of this paper, the term `internal deployment' refers to all possible uses of internally deployed AI models and systems, unless otherwise specified.} This use is often referred to as \textbf{autonomous AI Research \& Development}, or simply \textbf{autonomous AI R\&D}.\footnote{For clarity, AI R\&D is used to refer to both AI R\&D as a capability (i.e., an AI model or system's ability to accelerate and/or automate AI R\&D) and as a potential internal application of AI models or systems (and the relevant threat model).} While it is not currently feasible to fully automate the AI R\&D process,\footnote{\emph{See}, for instance, \href{https://www.anthropic.com/transparency/model-report}{Anthropic}, 2025 (``Claude Opus 4.5 could not fully automate an entry-level, remote-only research role at Anthropic''); \href{https://arxiv.org/pdf/2601.03267}{GPT-5 System Card}, 2025, at 41-42; \href{https://storage.googleapis.com/deepmind-media/Model-Cards/Gemini-3-Pro-Model-Card.pdf}{Gemini 3 Pro Model Card}, 2025, at 8, and \href{https://storage.googleapis.com/deepmind-media/gemini/gemini_3_pro_fsf_report.pdf}{Gemini 3 Pro Frontier Safety Framework Report}, at 14-15. \emph{See also} \href{https://metr.org/AI_R_D_Evaluation_Report.pdf}{Wijk et al}., 2024, at 21 (``a significant gap remains compared to the top human performance in most environments''); \href{https://metr.org/blog/2025-03-19-measuring-ai-ability-to-complete-long-tasks/}{METR}, 2026 (noting that ``the best AI agents are not currently able to carry out substantive projects by themselves or directly substitute for human labor'' and observing, in particular, that current AI models only ``succeed \textless{}10\% of the time on tasks taking more than around 4 hours'').} AI models and systems are becoming increasingly capable of accomplishing longer and more complex AI R\&D tasks.\footnote{\emph{See}, for instance, \href{https://metr.org/blog/2025-03-19-measuring-ai-ability-to-complete-long-tasks/}{METR}, 2026 (tracking the task-completion time horizons for the most advanced AI models publicly available); \href{https://arxiv.org/abs/2503.14499}{Kwa et al., 2025}, at 23 (noting that ``50\% task completion time horizon on our tasks has been growing exponentially from 2019--2025 with a doubling time of approximately seven months'' and that ``an 80\% confidence interval for the release date of AI that can complete 1-month long software tasks spans from late 2028 to early 2031'').} A frontier AI company recently stated in their system card that ``it is plausible that models equipped with highly effective scaffolding may not be very far away from ... AI R\&D-4 threshold,'' which captures the ability of an AI model to ``fully automate the work of an entry-level, remote-only researcher'' at that frontier AI company.\footnote{\href{https://assets.anthropic.com/m/64823ba7485345a7/Claude-Opus-4-5-System-Card.pdf}{Claude Opus 4.5 System Card}, 2025, at 13-14.} Existing progress in autonomous AI R\&D capabilities has recently spurred another frontier AI company to announce that they could achieve ``an automated AI research intern by September of 2026 \ldots{} and a true automated AI researcher by March of 2028.''\footnote{\emph{See} \href{https://x.com/sama/status/1983584366547829073}{Altman}, 2025; \href{https://techcrunch.com/2025/10/28/sam-altman-says-openai-will-have-a-legitimate-ai-researcher-by-2028/}{TechCrunch}, 2025. \emph{See also} \href{https://openai.com/index/ai-progress-and-recommendations/}{OpenAI}, 2025 (``we get closer to systems capable of recursive self-improvement.'')}

Legal and policy frameworks on both sides of the Atlantic are paying increasing attention to internal deployment and autonomous AI R\&D,\footnote{\emph{See} Section 22757.12.(a)(10), \href{https://leginfo.legislature.ca.gov/faces/billTextClient.xhtml?bill_id=202520260SB53}{California Senate Bill 53}, 2025; Section 3.3(e)(i), \href{https://bidenwhitehouse.archives.gov/briefing-room/presidential-actions/2024/10/24/memorandum-on-advancing-the-united-states-leadership-in-artificial-intelligence-harnessing-artificial-intelligence-to-fulfill-national-security-objectives-and-fostering-the-safety-security/}{National Security Memorandum}, 2024 (footnotes 18 and 19 below).} spurring the question as to whether and to what extent the AI Act covers internal deployment. For example, in the United States, California Senate Bill 53 requires ``large frontier developer[s]'' to ``[a]ssess[] and manag[e] catastrophic risk resulting from the \emph{internal use of} [their] \emph{frontier models}.''\footnote{Section 22757.12.(a)(10), \href{https://leginfo.legislature.ca.gov/faces/billTextClient.xhtml?bill_id=202520260SB53}{California Senate Bill 53}, 2025 (emphasis added).} Similarly, President Biden's National Security Memorandum required the U.S. Department of Commerce to concentrate on the testing of a series of AI capabilities that included the capability to ``automate development and deployment of other models with such capabilities.''\footnote{Section 3.3(e)(i), \href{https://bidenwhitehouse.archives.gov/briefing-room/presidential-actions/2024/10/24/memorandum-on-advancing-the-united-states-leadership-in-artificial-intelligence-harnessing-artificial-intelligence-to-fulfill-national-security-objectives-and-fostering-the-safety-security/}{National Security Memorandum}, 2024 (Memorandum on Advancing the United States' Leadership in Artificial Intelligence; Harnessing Artificial Intelligence to Fulfill National Security Objectives; and Fostering the Safety, Security, and Trustworthiness of Artificial Intelligence) (emphasis added).} At the same time, frontier AI companies have included dedicated thresholds for automated AI R\&D capabilities within their frontier safety policies.\footnote{\emph{See} \href{https://cdn.openai.com/pdf/18a02b5d-6b67-4cec-ab64-68cdfbddebcd/preparedness-framework-v2.pdf}{OpenAI}, 2025; \href{https://www-cdn.anthropic.com/872c653b2d0501d6ab44cf87f43e1dc4853e4d37.pdf}{Anthropic}, 2025; \href{https://storage.googleapis.com/deepmind-media/DeepMind.com/Blog/strengthening-our-frontier-safety-framework/frontier-safety-framework_3.pdf}{Google DeepMind}, 2025; \href{https://metr.org/common-elements.pdf}{METR}, 2025.} Through these internal policies, frontier AI companies have committed to, for example, evaluating ``the ability of an AI system to accelerate AI research, including to increase the system's own capability,''\footnote{\href{https://cdn.openai.com/pdf/18a02b5d-6b67-4cec-ab64-68cdfbddebcd/preparedness-framework-v2.pdf}{OpenAI}, 2025, at 6. \emph{See}, e.g., \href{https://cdn.openai.com/gpt-5-system-card.pdf}{GPT-5 System Card}, 2025, at 41.} ``[t]he ability to fully automate the work of an entry-level, remote-only researcher'' and ``to cause dramatic acceleration in the rate of effective scaling,''\footnote{\href{https://www-cdn.anthropic.com/872c653b2d0501d6ab44cf87f43e1dc4853e4d37.pdf}{Anthropic}, 2025, at 4. \emph{See}, e.g., \href{https://assets.anthropic.com/m/64823ba7485345a7/Claude-Opus-4-5-System-Card.pdf}{Claude Opus 4.5 System Card}, 2025, at 13-14; \href{https://www-cdn.anthropic.com/0dd865075ad3132672ee0ab40b05a53f14cf5288.pdf}{Claude Opus 4.6 System Card}, 2026, at 12-13, 181-194} and whether this capability can ``result[] in AI progress substantially accelerating from historical rates.''\footnote{\href{https://storage.googleapis.com/deepmind-media/DeepMind.com/Blog/strengthening-our-frontier-safety-framework/frontier-safety-framework_3.pdf}{Google DeepMind}, 2025, at 13. In addition to an ``ML R\&D acceleration'' capability threshold, Google DeepMind also has an ``ML R\&D automation'' capability threshold, capturing whether an AI model or system can ``fully automate the work of any team of researchers at Google focused on improving AI capabilities, with approximately comparable all-inclusive costs'' (\href{https://storage.googleapis.com/deepmind-media/DeepMind.com/Blog/strengthening-our-frontier-safety-framework/frontier-safety-framework_3.pdf}{Google DeepMind}, 2025, at 14). \emph{See}, e.g., \href{https://storage.googleapis.com/deepmind-media/Model-Cards/Gemini-3-Pro-Model-Card.pdf}{Gemini 3 Pro Model Card}, 2025, at 8.}

The Safety and Security Chapter of the Code of Practice (COP) for General-Purpose AI (GPAI) Models\footnote{For clarity, all references to the COP in this memorandum refer to its Safety and Security Chapter.} hints at internal deployment for AI R\&D purposes\footnote{\emph{See} Measure 7.1, \href{https://digital-strategy.ec.europa.eu/en/policies/contents-code-gpai}{COP} (``Signatories will provide in the Model Report: \ldots{} a description of how the model has been \emph{used} and is expected to be used, including its use \emph{in the development, oversight, and/or evaluation of models}) (emphasis added); Measure 7.3, \href{https://digital-strategy.ec.europa.eu/en/policies/contents-code-gpai}{COP} (``Signatories will provide in the Model Report: \ldots{} (4) a high-level description of: (a) the techniques and assets they intend to use to further develop the model over the next six months, including \emph{through the use of other AI models and/or AI systems}\ldots{}'') (emphasis added).} and includes the ``capabilities to automate AI research and development'' (i.e., autonomous AI R\&D) within the ``sources of systemic risks,''\footnote{Appendix 1.3, \href{https://digital-strategy.ec.europa.eu/en/policies/contents-code-gpai}{COP}. \emph{See also} Recital 110, \href{https://eur-lex.europa.eu/legal-content/EN/TXT/PDF/?uri=OJ:L_202401689}{AI Act} (mentioning the ``risks from models ... training other models'' in the context of systemic risk from GPAI models).} making the question around the application of the AI Act to internally deployed AI models and systems all the more urgent and consequential.\footnote{If internal GPAI models are covered by the AI Act, signatories will have to evaluate these internal GPAI models for the ``capabilities to automate AI research and development'' (Appendix 1.3.1, \href{https://digital-strategy.ec.europa.eu/en/policies/contents-code-gpai}{COP}). This may be particularly consequential because the AI models and systems used for AI R\&D purposes are internal, rather than external. In this respect, it is worth observing that some frontier AI companies already evaluate \emph{external} AI models and systems for AI R\&D capabilities (\emph{see} footnotes 20-23 above).} Nonetheless, compared to other legal frameworks mentioning internal deployment expressly (for instance, California Senate Bill 53), the AI Act does not clearly state that internal AI models or systems fall within its scope.

This memorandum examines both sides of the equation: it considers interpretative options to deem internal deployment as included within the scope of the AI Act (\S{}2), as well as possible counterarguments and exceptions (\S{}3).\footnote{In general, the question around the application of the AI Act to internal deployment should not be confused with the obligations AI providers may have for \emph{covered} AI models and systems \emph{before} \emph{their public deployment}. Examples of the latter are Article 55 of the AI Act and Recital (a) of the COP, which require providers to ``take appropriate measures'' to ``assess and mitigate'' ``systemic risks'' by ``taking appropriate measures \emph{along the entire model lifecycle} (including during development that occurs \emph{before \ldots{} a model has been placed on the market})'' (emphasis added). Similarly, Measures 1.2 and 4.2 of the COP require providers to undertake these assessments ``\emph{[a]long the entire model lifecycle}'' and ``only proceed with the \emph{development}, the making available on the market, and/or the use of the model, if the systemic risks stemming from the model are determined to be acceptable'' (\href{https://digital-strategy.ec.europa.eu/en/policies/contents-code-gpai}{COP}; emphasis added).} This memorandum elaborates on, and stress-tests, Articles 2(1)(a)-(c), 2(6), and 2(8) of the AI Act, which offer potential interpretative pathways for supporting or opposing the inclusion of internal deployment within the scope of the AI Act. Finally, the conclusion of this memorandum (\S{}4) puts forward a proposal for how all these, potentially conflicting, provisions of the AI Act can be interpreted systematically---and, ultimately, all brought to coherence. As the conclusion of this memorandum illustrates, once broken down to their most plausible meaning and interpreted in conjunction with other relevant provisions,\footnote{\emph{See}, in particular, Articles 3(1), 3(3), 3(4), 3(9), 3(10), 3(11), 3(12), 3(63) and Recitals 12, 13, 21, 25, 97, 109, and 110, \href{https://eur-lex.europa.eu/legal-content/EN/TXT/PDF/?uri=OJ:L_202401689}{AI Act}.} Articles 2(1), 2(6), and 2(8) of the AI Act are, in fact, complementary to each other (\S{}4 and Figure 1 above, at 1). In summary, the internal deployment of AI systems or GPAI models integrated into AI systems could trigger the application of the AI Act under Article 2(1)(a)-(c), unless an internal AI model or system is specifically developed and deployed for the sole purpose of scientific R\&D under Article 2(6). Pre-internal-deployment research, testing and development activity arguably falls outside the scope of the AI Act under Article 2(8).

\subsection*{2. Arguments Supporting the Application of the AI Act to Internal Deployment}
\addcontentsline{toc}{subsection}{2. Arguments Supporting the Application of the AI Act to Internal Deployment}

This section explores potential interpretations of the AI Act that support the inclusion of internal deployment within the scope of the Act. Specifically, \S{}2.1 examines the potential inclusion of \textbf{internal AI systems} within the scope of the AI Act. \S{}2.2 then discusses the potential inclusion of \textbf{internal GPAI models} within the scope of the AI Act.

The following \S{}3 will then lay out potential interpretations of the AI Act that support the exclusion of internal deployment from the scope of the Act, or at least restrict the AI Act's scope only to some AI models and systems deployed internally, as well as potential counterarguments.

\subsection*{2.1 Internal AI Systems}
\addcontentsline{toc}{subsection}{2.1 Internal AI Systems}

\S{}\S{}2.1.1-2.1.4 of this section concentrate on Article 2 of the AI Act, which defines the ``scope'' of the Act, and offer reasons why the deployment of \textbf{internal AI systems} could be considered as covered by the AI Act.\footnote{For clarity, \S{}\S{}21-2.4 discuss the potential application of the AI Act to internal \emph{AI systems}. \S{}3.1 below will discuss the potential application of the AI Act to internal \emph{GPAI models} and the internal AI systems integrating GPAI models, thus looping back to \S{}\S{}2.1-2.4.} Specifically:

\begin{itemize}
  \item \textbf{\S{}2.1.1} concentrates on the expression ``providers \ldots{} putting into service AI systems'' in Article 2(1)(a).
  \item \textbf{\S{}2.1.2} concentrates on the expression ``providers placing on the market \ldots{} AI systems'' in Article 2(1)(a).\footnote{\S{}3.1 below concentrates on the expression ``placing on the market \emph{general-purpose AI models}'' in Article 2(1)(a) (emphasis added).}
  \item \textbf{\S{}2.1.3} concentrates on the expression ``deployers of AI systems that have their place of establishment or are located within the Union'' in Article 2(1)(b).
  \item \textbf{\S{}2.1.4} concentrates on the expression ``output produced by the AI system is used in the Union'' in Article 2(1)(c).
\end{itemize}

\subsection*{2.1.1 Article 2(1)(a): ``Providers \ldots{} putting into service AI systems''}
\addcontentsline{toc}{subsection}{2.1.1 Article 2(1)(a): ``Providers \ldots{} putting into service AI systems''}

Article 2 of the AI Act defines the ``scope'' of the Act.\footnote{The title of Article 2 of the AI Act is, in fact, ``Scope'' (\href{https://eur-lex.europa.eu/legal-content/EN/TXT/PDF/?uri=OJ:L_202401689}{AI Act}).} Specifically, under Article 2(1)(a), the AI Act ``applies to'' ``providers \ldots{} putting into service AI systems'' in the Union.\footnote{Article 2(1), \href{https://eur-lex.europa.eu/legal-content/EN/TXT/PDF/?uri=OJ:L_202401689}{AI Act} (``This Regulation applies to: (a) providers placing on the market or putting into service AI systems or placing on the market general-purpose AI models in the Union, irrespective of whether those providers are established or located within the Union or in a third country'').} This section examines whether the expression ``providers \ldots{} putting into service AI systems''\footnote{Article 2(1)(a), \href{https://eur-lex.europa.eu/legal-content/EN/TXT/PDF/?uri=OJ:L_202401689}{AI Act}.} could include AI systems internally deployed by the providers developing them.

The AI Act offers definitions for all the components of this expression (``providers \ldots{} putting into service AI systems''\footnote{Article 2(1)(a), \href{https://eur-lex.europa.eu/legal-content/EN/TXT/PDF/?uri=OJ:L_202401689}{AI Act}.}). First, the term ``\textbf{providers}'' is defined by Article 3(3) as ``a natural or legal person, public authority, agency or other body that develops an AI system or a general-purpose AI model or that has an AI system or a general-purpose AI model developed and places it on the market or puts the AI system into service under its own name or trademark, whether for payment or free of charge.'' Importantly, both Article 2(1)(a) and Recital 21 clarify that the AI Act applies to providers of AI systems ``irrespective of whether those providers are established or located within the Union or in a third country.''\footnote{Article 2(1)(a), \href{https://eur-lex.europa.eu/legal-content/EN/TXT/PDF/?uri=OJ:L_202401689}{AI Act}. \emph{See also} Recital 21, \href{https://eur-lex.europa.eu/legal-content/EN/TXT/PDF/?uri=OJ:L_202401689}{AI Act} (``the rules established by this Regulation should apply to providers of AI systems in a non-discriminatory manner, \emph{irrespective of whether they are established within the Union or in a third country}, and to deployers of AI systems established within the Union'') (emphasis added). \emph{See}, for instance, \href{https://law-store.wolterskluwer.com/s/product/the-eu-artificial-intelligence-ai-act-a-commentary/01tPg000007gkK9IAI?srsltid=AfmBOooB8DDUCEJELRg4SkWxkoS853DEceZpZ3x659nmpw3IsLNZ6WuM}{Van Eecke and Regenhardt}, 2024, at 35; \href{https://doi.org/10.1017/glj.2023.108}{Almada and Radu}, 2024, at 656; \href{http://Czerniawski}{Czerniawski}, 2024, at 4.} Therefore, the term providers encompasses \textbf{both EU-based and foreign AI developers}.

Second, ``\textbf{putting into service}'' is defined by Article 3(11) as ``the \emph{supply} of an AI system for first use directly to the deployer or \emph{for own use} \emph{in the Union} for its intended purpose.''\footnote{\emph{See} the `Blue Guide' on the implementation of EU product rules 2022, which clarifies that ``putting into service'' ``results in the scope of Union harmonisation legislation being extended beyond the moment of making available of a product'' (Section 2.6, \href{https://eur-lex.europa.eu/legal-content/EN/TXT/PDF/?uri=CELEX:52022XC0629(04)}{Blue Guide}). Specifically, ``[p]utting into service takes place at the moment of first use within the Union by the end user for the purposes for which it was intended'' (Section 2.6, \href{https://eur-lex.europa.eu/legal-content/EN/TXT/PDF/?uri=CELEX:52022XC0629(04)}{Blue Guide}).} While the AI Act does not define ``supply,'' it is possible to infer that its meaning corresponds to ``making available.''\footnote{\emph{See}, for instance: (i) under the `Blue Guide,' ``[s]upplying a product is only considered as \emph{making available} on the Union market, when the product is intended for end use on the Union market'' (Section 2.2, \href{https://eur-lex.europa.eu/legal-content/EN/TXT/PDF/?uri=CELEX:52022XC0629(04)}{Blue Guide}; emphasis added); and (ii) the definition of ``making available on the market'' is ``supply'' in Article 3(1), \href{https://eur-lex.europa.eu/legal-content/EN/TXT/PDF/?uri=CELEX:32019R1020}{Regulation (EU) 2019/1020 of June 20, 2019}, Article 2(1), \href{https://eur-lex.europa.eu/legal-content/EN/TXT/PDF/?uri=CELEX:32008R0765}{Regulation (EC) 765/2008 of July 9, 2008}, Article 3(6), \href{https://eur-lex.europa.eu/legal-content/EN/TXT/PDF/?uri=CELEX:32023R0988}{Regulation (EU) 2023/988 of May 10, 2023}, and Article 2(27), \href{https://eur-lex.europa.eu/legal-content/EN/TXT/PDF/?uri=CELEX:32017R0745}{Regulation (EU) 2017/745 of April 5, 2017}.} Article 3(3) clarifies that ``putting into service'' an AI system can be ``for payment or free of charge\emph{.}''\footnote{The AI Act also defines ``intended purpose.'' Under Article 3(12), ``intended purpose'' means ``the use for which an AI system is intended by the provider, including the specific context and conditions of use, as specified in the information supplied by the provider in the instructions for use, promotional or sales materials and statements, as well as in the technical documentation'' (\href{https://eur-lex.europa.eu/legal-content/EN/TXT/PDF/?uri=OJ:L_202401689}{AI Act}). Therefore, based on this definition, if a provider intended to deploy an AI system internally, arguably that is the system's intended purpose.} Therefore, given these reference points, ``putting into service'' can be interpreted as including a provider's act of \textbf{making available an AI system for the provider's own use in the Union}.\footnote{It remains unclear how ``in the Union'' will be interpreted. \emph{See} Article 3(11), \href{https://eur-lex.europa.eu/legal-content/EN/TXT/PDF/?uri=OJ:L_202401689}{AI Act}.}

Third, ``\textbf{AI system}'' is defined by Article 3(1) as ``a machine-based system that is designed to operate with varying levels of autonomy and that may exhibit adaptiveness after deployment, and that, for explicit or implicit objectives, infers, from the input it receives, how to generate outputs such as predictions, content, recommendations, or decisions that can influence physical or virtual environments.''\footnote{\emph{See also} Commission Guidelines on the definition of an AI system established by Regulation (EU) 2024/1689 (AI Act) (\href{https://digital-strategy.ec.europa.eu/en/library/commission-publishes-guidelines-ai-system-definition-facilitate-first-ai-acts-rules-application}{European Commission}, 2025).} Importantly, the definition of AI systems does not include any restrictions on who, amongst internal and external users, can give ``input'' to an AI system and/or use its ``outputs.''\footnote{Article 3(1), \href{https://eur-lex.europa.eu/legal-content/EN/TXT/PDF/?uri=OJ:L_202401689}{AI Act}.} In other words, the definition of ``AI system'' in Article 3(1) does not differentiate between external and internal AI systems. Therefore, at least in theory, the definition of an AI system could include \textbf{both external and internal AI systems}.

By collating these various definitions, the expression ``[p]roviders \ldots{} putting into service AI systems''\footnote{Article 2(1)(a), \href{https://eur-lex.europa.eu/legal-content/EN/TXT/PDF/?uri=OJ:L_202401689}{AI Act}.} can be reasonably interpreted to include \textbf{EU-based or foreign AI developers making available an internal AI system for their own use in the Union}. In other words, Article 2(1)(a) could arguably be interpreted to include internal AI systems within the scope of the AI Act. Indeed, absent any express exclusion of internal use from the scope of the AI Act (as identified in Article 2 of the AI Act),\footnote{\emph{See also}, for reference, the Glossary in the \href{https://digital-strategy.ec.europa.eu/en/policies/contents-code-gpai}{COP}, defining ``use (of a model)'' simply as ``use of the model by the Signatory or other actors.''} ``[p]roviders \ldots{} putting into service AI systems'' could arguably be interpreted to include AI systems that EU-based or foreign AI developers deploy for internal use in the Union.\footnote{For instance, AI developers making internal AI systems accessible to staff based in their EU offices.}

\subsection*{2.1.2. Article 2(1)(a): ``Providers placing on the market \ldots{} AI systems''}
\addcontentsline{toc}{subsection}{2.1.2. Article 2(1)(a): ``Providers placing on the market \ldots{} AI systems''}

\S{}2.1.1 above started examining the scope of the AI Act, concentrating on Article 2(1)(a). As mentioned above, under Article 2(1)(a), the AI Act ``applies to'' ``[p]roviders \ldots{} putting into service AI systems.''\footnote{Article 2(1), \href{https://eur-lex.europa.eu/legal-content/EN/TXT/PDF/?uri=OJ:L_202401689}{AI Act} (``This Regulation applies to: (a) providers placing on the market or putting into service AI systems or placing on the market general-purpose AI models in the Union, irrespective of whether those providers are established or located within the Union or in a third country'').} Under the same Article, the AI Act also applies to ``[p]roviders placing on the market \ldots{} AI systems.''\footnote{Article 2(1)(a), \href{https://eur-lex.europa.eu/legal-content/EN/TXT/PDF/?uri=OJ:L_202401689}{AI Act}.} Building on the considerations on ``providers'' and ``AI system'' developed in \S{}2.1.1 above, this section explores whether ``placing on the market''\footnote{Article 2(1)(a), \href{https://eur-lex.europa.eu/legal-content/EN/TXT/PDF/?uri=OJ:L_202401689}{AI Act}.} could potentially refer to internally deployed AI systems. In doing so, this section examines Article 3(9)-(10) and assesses whether these Articles support a broad interpretation of ``placing on the market'' that includes internal AI systems.

Article 3(9) defines ``\textbf{placing on the market}'' as ``the \emph{first making available} of an AI system or a general-purpose AI model \emph{on the Union market}.''\footnote{Article 3(9), \href{https://eur-lex.europa.eu/legal-content/EN/TXT/PDF/?uri=OJ:L_202401689}{AI Act} (emphasis added).} Article 3(10) further defines ``making available on the market'' as ``the \emph{supply} of an AI system or a general-purpose AI model for distribution or \emph{use} \emph{on the Union market} \emph{in the course of a commercial activity}, whether in return for payment or \emph{free of charge.}''\footnote{Article 3(10), \href{https://eur-lex.europa.eu/legal-content/EN/TXT/PDF/?uri=OJ:L_202401689}{AI Act} (emphasis added).} As mentioned in \S{}2.1.1 above, ``supply'' can be interpreted as ``making available.'' The AI Act does not define ``use,'' which other relevant EU frameworks define broadly.\footnote{\emph{See}, for instance, Section 2.2, \href{https://eur-lex.europa.eu/legal-content/EN/TXT/PDF/?uri=CELEX:52022XC0629(04)}{Blue Guide} (defining ``use'' as ``the intended purpose of the product as defined by the manufacturer under conditions which can be reasonably foreseen''); Section 2.3, Commission Guidelines on prohibited AI practices established by Regulation (EU) 2024/1689 (AI Act) (``While the `use' of an AI system is not explicitly defined in the AI Act, it should be understood in a broad manner to cover the use or deployment of the system at \emph{any moment of its lifecycle} \emph{after having been} placed on the market or \emph{put into service}'') (\href{https://digital-strategy.ec.europa.eu/en/library/commission-publishes-guidelines-prohibited-artificial-intelligence-ai-practices-defined-ai-act}{European Commission}, 2025, at 5; emphasis added). Therefore, to distil the meaning of ``use,'' it is possible to draw inspiration from these frameworks.} Neither does the AI Act define ``in the course of a commercial activity.''\footnote{Article 3(10), \href{https://eur-lex.europa.eu/legal-content/EN/TXT/PDF/?uri=OJ:L_202401689}{AI Act}.} However, to distil its meaning it is possible to draw inspiration from the European Commission's Blue Guide on the implementation of EU product rules,\footnote{\emph{See} the \href{https://eur-lex.europa.eu/legal-content/EN/TXT/PDF/?uri=CELEX:52022XC0629(04)}{Blue Guide}. For clarity, the Blue Guide may be an appropriate reference because ``[t]he AI Act is rooted in classic EU product safety law'' (\href{https://scholarship.law.umn.edu/cgi/viewcontent.cgi?article=1576&context=mjlst}{Boine and Rolnick}, 2025, at 84). \emph{See also} \href{https://www.researchgate.net/publication/380319968_The_European_Union's_Regulatory_Challenge_Conceptualizing_Purpose_in_Artificial_Intelligence}{Lanamäki et al}., 2024, at 5 (discussing how the first proposal of AI Act by the European Commission in 2021 was based on EU product safety principles).} as well as other EU legal frameworks, such as for instance Regulation (EU) 2023/1115.\footnote{\emph{See} Article 2(19), \href{https://eur-lex.europa.eu/legal-content/EN/TXT/HTML/?uri=CELEX:02023R1115-20241226}{Regulation (EU) 2023/1115 of May 31, 2023}.} Within the first framework, ``[c]ommercial activity is understood as providing goods \emph{in a business related context}.''\footnote{Section 2.2, \href{https://eur-lex.europa.eu/legal-content/EN/TXT/PDF/?uri=CELEX:52022XC0629(04)}{Blue Guide} (emphasis added). It also clarifies that ``[n]on-profit organisations may be considered as carrying out commercial activities if they operate in such a context. This can only be appreciated on a case by case basis taking into account the regularity of the supplies, the characteristics of the product, the intentions of the supplier, etc.''} The second defines ``in the course of a commercial activity'' as ``\emph{for use in the business of the operator or trader itself.}''\footnote{\emph{See} Article 2(19), \href{https://eur-lex.europa.eu/legal-content/EN/TXT/HTML/?uri=CELEX:02023R1115-20241226}{Regulation (EU) 2023/1115 of May 31, 2023} (emphasis added).} In other words, ``in the course of a commercial activity''\footnote{Article 3(10), \href{https://eur-lex.europa.eu/legal-content/EN/TXT/PDF/?uri=OJ:L_202401689}{AI Act}.} does not necessarily mean `for sale' (especially considering that Articles 3(3) and 3(10) clarify that the ``supply'' can be ``free of charge,'' which is not the case within other regulatory frameworks\footnote{Other EU legal frameworks take a different approach. \emph{See}, for instance, Recital 14, \href{https://eur-lex.europa.eu/legal-content/EN/TXT/PDF/?uri=OJ:L_202402853}{Directive (EU) 2024/2853 of October 23, 2024} (``the supply of free and open-source software by non-profit organisations should not be considered as taking place in a business-related context, unless such supply occurs in the course of a commercial activity. However, where software is supplied in \emph{exchange for a price}, or for \emph{personal data} used other than exclusively for improving the security, compatibility or interoperability of the software, and is therefore supplied in the course of a commercial activity, this Directive should apply;'' emphasis added).}). Goods can be used in the business of the operator or trader itself for many different purposes other than sale, including for improving internal business procedures or increasing efficiency in ways that enhance a company's competitiveness. Therefore, based on the definition contained in Article 3(9), as interpreted through Article 3(10), ``placing on the market'' in Article 2(1)(a) could be interpreted as \textbf{making available an AI system for use on the Union market}\footnote{It remains unclear how ``on the Union market'' will be interpreted. \emph{See} Article 3(9), \href{https://eur-lex.europa.eu/legal-content/EN/TXT/PDF/?uri=OJ:L_202401689}{AI Act}.} \textbf{in a business-related context, including for use in the business of the provider itself}. This seems confirmed, among other things, by the European Commission's Guidelines on the scope of the obligations for GPAI models, which, in listing ``examples of placing on the market of general-purpose AI models,'' include a GPAI model being ``used for internal processes.''\footnote{\emph{See} Section 3.1.2., Guidelines on the scope of the obligations for general-purpose AI models established by Regulation (EU) 2024/1689 (AI Act) (\href{https://digital-strategy.ec.europa.eu/en/library/guidelines-scope-obligations-providers-general-purpose-ai-models-under-ai-act}{European Commission}, 2025, at 18).} Hence, Article 2(1)(a) could arguably support the inclusion of internal AI system use within the scope of the AI Act.

While this interpretation closely reflects the content of Article 3(9)-(10), some doubts remain as to whether this interpretation sufficiently accounts for the natural meaning of ``placing on the market.''\footnote{Article 2(1)(a), \href{https://eur-lex.europa.eu/legal-content/EN/TXT/PDF/?uri=OJ:L_202401689}{AI Act}. These doubts could also potentially be corroborated by the fact that Article 3(11) (defining ``putting into service'') refers to ``use \emph{in the Union}'' (emphasis added), whereas Article 3(10) (defining ``making available on the market'') refers to ``use \emph{on the Union market}'' (emphasis added).} Ultimately, if this interpretation of ``placing on the market'' were to prevail, the only difference with ``putting into service'' would be that the latter could refer to any and all contexts, including ones that are not business-related, whereas ``placing on the market'' would cover only business-related contexts.

In conclusion, while the common meaning of ``placing of the market'' does not immediately present as encompassing the internal deployment of AI systems, it could arguably be interpreted in a sufficiently extensive way so as to cover an EU or foreign provider's act of making an AI system available on the EU market for access and use by its EU staff. Indeed, most internal uses of AI systems by providers plausibly fall within a business-related context. For instance, as mentioned in \S{}1 above, an internal system could support a provider's staff by effectively outsourcing portions of existing or new workstreams; and, in a not-so-distant future, internal AI models and systems might start automating work.\footnote{\emph{See} footnote 14 above.}

\subsection*{2.1.3 Article 2(1)(b): ``Deployers of AI systems that have their place of establishment or are located within the Union''}
\addcontentsline{toc}{subsection}{2.1.3 Article 2(1)(b): ``Deployers of AI systems that have their place of establishment or are located within the Union''}

Under Article 2(1)(b), the AI Act ``applies to'' ``[d]eployers of AI systems that have their place of establishment or are located within the Union.'' This section examines whether and to what extent Article 2(1)(b) could potentially include AI systems deployed within the AI providers developing them.

Article 3(4) and Recital 13 define ``\textbf{deployer}'' as ``a natural or legal person, public authority, agency or other body \emph{using} an AI system under its authority.''\footnote{Article 3(4) and Recital 13, \href{https://eur-lex.europa.eu/legal-content/EN/TXT/PDF/?uri=OJ:L_202401689}{AI Act} (emphasis added).} Indirectly, this means that `deploying an AI system' is generically defined as `using an AI system.' Neither Article 3(4) nor Recital 13 restrict in any way the type of `use' that a natural or legal person can make of AI systems to qualify as a deployer. Therefore, under Article 3(4) and Recital 13, \textbf{a provider and/or a provider's staff using AI systems internally could qualify as a} ``\textbf{deployer.}''

Importantly, Article 2(1)(b) limits the geographic area of relevance to deployers (i.e., users) that ``have their place of establishment or are located within the Union.''\footnote{Article 2(1)(b), \href{https://eur-lex.europa.eu/legal-content/EN/TXT/PDF/?uri=OJ:L_202401689}{AI Act}.} By contrast, Article 2(1)(b) does not set any jurisdictional limits as to where the AI system is developed. In other words, while the deployer of an AI system must be established or located in the EU under Article 2(1)(b), the provider of that system can be established or located elsewhere.\footnote{\emph{See} \S{}2.1.1 above. Therefore, for example, members of the technical staff of a foreign AI provider who are located in the EU could potentially qualify as `deployer' if they use an internal coding agent developed by the AI provider overseas and then made available to staff globally to support an existing workstream.} Article 2(1)(b) is geographically constrained as to system deployment (i.e., use), but jurisdiction-agnostic as to its development.

In conclusion, the definition of `deployment' as, generically, the `use' of an AI system could be interpreted to encompass \textbf{any and all potential internal uses that a EU-based AI provider or the EU-based staff of a foreign provider could make of an AI system}.\footnote{\emph{See} some examples of internal use in \S{}1 above.}  This conclusion appears to be supported by the text of the COP, which utilizes the same verb (`using') to describe one possible form of internal deployment---autonomous AI R\&D.\footnote{Measure 7.1, \href{https://digital-strategy.ec.europa.eu/en/policies/contents-code-gpai}{COP} (``Signatories will provide in the Model Report: \ldots{} a description of how the model has been \emph{used} and is expected to be used, including its \emph{use in the development, oversight, and/or evaluation of models}; \ldots{} a description of the model versions that are going to be made or are currently made available on the market and/or used, including differences in systemic risk mitigations and systemic risks'') (emphasis added).} Specifically, Measure 7.1 acknowledges that an AI system can be ``\emph{used}'' ``in the \emph{development}, oversight, and/or evaluation of models.''\footnote{Measure 7.1, \href{https://digital-strategy.ec.europa.eu/en/policies/contents-code-gpai}{COP}.} In other words, an AI system can be `used' (i.e., deployed) for autonomous AI R\&D, which is one possible use of internally deployed systems.

\subsection*{2.1.4 Article 2(1)(c): ``output produced by the AI system is used in the Union''}
\addcontentsline{toc}{subsection}{2.1.4 Article 2(1)(c): ``output produced by the AI system is used in the Union''}

Under Article 2(1)(c), the AI Act applies to ``providers and deployers of AI systems that have their place of establishment or are located in a \emph{third country}, where the output produced by the AI system is used in the Union.''\footnote{Article 2(1)(c), \href{https://eur-lex.europa.eu/legal-content/EN/TXT/PDF/?uri=OJ:L_202401689}{AI Act} (emphasis added).} Building on the conclusions reached in the previous \S{}\S{}2.1.1-2.1.3 regarding the terms `provider,' `deployer,' and `AI system,' this section examines whether ``the output produced by the AI system is used in the Union'' could be interpreted to support the inclusion of internally deployed AI systems within the scope of the AI Act.

According to Recital 12, ``outputs generated by the AI system reflect different functions performed by AI systems and include \emph{predictions, content, recommendations or decisions.}''\footnote{Recital 12, \href{https://eur-lex.europa.eu/legal-content/EN/TXT/PDF/?uri=OJ:L_202401689}{AI Act} (emphasis added).} Neither this description of ``output,'' nor the definition of ``AI system''\footnote{\emph{See} \S{}2.1.1 above.} offer any element to reasonably exclude internal AI systems. Like external systems, internal systems can also output ``predictions, content, recommendations or decisions.''\footnote{Recital 12, \href{https://eur-lex.europa.eu/legal-content/EN/TXT/PDF/?uri=OJ:L_202401689}{AI Act}.} Therefore, it seems reasonable to conclude that ``output produced by the AI system'' could also include the \textbf{output produced by internal AI systems} deployed by a provider.

Compared to Article 2(1)(b),\footnote{\emph{See} \S{}2.1.3 above.} which concerns deployers established or located within the EU, Article 2(1)(c) concerns ``providers and deployers \ldots{} that have their place of establishment or are located in a \emph{third country}.''\footnote{Article 2(1)(c), \href{https://eur-lex.europa.eu/legal-content/EN/TXT/PDF/?uri=OJ:L_202401689}{AI Act} (emphasis added).} This means that, if ``output produced by the AI system'' were interpreted to include the outputs of internal AI systems, Article 2(1)(c) would cover foreign AI providers developing and deploying an internal AI system outside of the EU, and sharing the output within the EU. This could occur, for example, in cases in which: (i) a foreign provider develops a highly capable AI system outside of the EU; (ii) the same foreign provider deploys internally that AI system outside of the EU to accelerate or automate AI R\&D; (iii) once deployed, the provider shares the research output generated by this internal AI R\&D system  with staff located in the EU; and (iv) this research output is used by the provider's staff within the EU.

\subsection*{2.2 Internal GPAI Models}
\addcontentsline{toc}{subsection}{2.2 Internal GPAI Models}

\S{}2.1 examined potential arguments supporting the inclusion of internal \emph{AI systems} within the scope of the AI Act. \S{}2.2 turns to internal \emph{GPAI models} and discusses whether internal GPAI models could be covered by the AI Act, focusing on two pathways: (i) the placement on the market of internal GPAI models via their integration into internal AI systems, pursuant to Article 2(1)(a) and Recital 97 of the AI Act; and (ii) an extensive interpretation of the expression ``placing on the market'' in Article 2(1)(a) of the AI Act as `making a GPAI model available for use on the Union market in a business-related context, including for use in the business of the provider itself' (\S{}2.1.2).

Under Article 2(1)(a), the AI Act ``applies to'' ``providers placing on the market or putting into service AI systems \emph{or placing on the market general-purpose AI models in the Union.}''\footnote{Article 2(1)(a), \href{https://eur-lex.europa.eu/legal-content/EN/TXT/PDF/?uri=OJ:L_202401689}{AI Act} (emphasis added).} Based on a summary reading, this Article could be interpreted to exclude the application of the AI Act to internal GPAI models. Specifically, Article 2(1)(a) could be read to indicate that: (i) while AI systems fall within the scope of the AI Act once they are ``put into service,'' the AI Act applies to GPAI models only after they are ``placed on the market;''\footnote{Article 2(1)(a), \href{https://eur-lex.europa.eu/legal-content/EN/TXT/PDF/?uri=OJ:L_202401689}{AI Act}.} and, therefore, (ii) if deploying (i.e., ``putting into service'') an AI system internally could potentially trigger the application of the AI Act (as argued in \S{}2.1 above), the same is not true about GPAI models (which are relevant only once they are ``placed on the market''). In other words, since Article 2(1)(a) does not use the expression ``putting into service'' with direct reference to GPAI models, deploying (i.e., ``putting into service'') a GPAI model internally would not trigger the application of the AI Act in the same way that deploying (i.e., ``putting into service'') an internal AI system does.\footnote{For clarity, this argument would refer only to internal GPAI models (rather than AI systems embedding GPAI models). AI systems integrating GPAI models could remain covered through Article 2(1)(a)-(c). \emph{See} \S{}\S{}2.1.1-2.1.4 above; \emph{see also} the analysis of Recital 97 below in \S{}3.1.} To support this interpretation, it would be possible to rely on Article 3(63), which excludes from the definition of GPAI model ``AI models that are used for research, development or prototyping activities \emph{before they are placed on the market},''\footnote{Article 3(63), \href{https://eur-lex.europa.eu/legal-content/EN/TXT/PDF/?uri=OJ:L_202401689}{AI Act} (```general-purpose AI model' means an AI model, including where such an AI model is trained with a large amount of data using self-supervision at scale, that displays significant generality and is capable of competently performing a wide range of distinct tasks regardless of the way the model is placed on the market and that can be integrated into a variety of downstream systems or applications, \emph{except AI models that are used for research, development or prototyping activities before they are placed on the market}'') (emphasis added).} and Recital 97. This latter Recital specifies that ``the obligations for the providers of general-purpose AI models should apply \emph{once} the general-purpose AI models are \emph{placed on the market},'' and that ``[t]he definition [of GPAI model] should \emph{not cover AI models used before their placing on the market for the sole purpose of research, development and prototyping activities}.''\footnote{Recital 97, \href{https://eur-lex.europa.eu/legal-content/EN/TXT/PDF/?uri=OJ:L_202401689}{AI Act} (emphasis added). Recital 97 also clarifies that ``[t]he obligations laid down for models should in any case not apply when an own model is used for purely internal processes that are not essential for providing a product or a service to third parties and the rights of natural persons are not affected.''}

Despite these arguments, there are reasons why internal GPAI models could still fall within the scope of the AI Act. First, as mentioned in \S{}2.1.2 above, ``placing on the market'' a GPAI model\footnote{Article 2(1)(a), \href{https://eur-lex.europa.eu/legal-content/EN/TXT/PDF/?uri=OJ:L_202401689}{AI Act}.} could potentially be interpreted extensively as `making a GPAI model available for use on the Union market in a business-related context, including for use in the business of the provider itself.' If ``placing on the market'' were to be interpreted in this extensive way, the internal deployment of GPAI models could still be subject to the AI Act under Article 2(1)(a).\footnote{As mentioned in \S{}2.1.2 above, this interpretation may risk eroding most of the difference between ``placing on the market'' and ``putting into service,'' and for this reason may not be pursued.}

Second, and \emph{more persuasively}: (i) it is plausible that, at least within frontier AI companies, internal deployment usually concerns GPAI models that are integrated into internal AI systems, and given all the necessary affordances and permissions; and therefore, (ii) after system integration (and once the relevant system is ``put into service'' internally\footnote{Article 2(1)(a), \href{https://eur-lex.europa.eu/legal-content/EN/TXT/PDF/?uri=OJ:L_202401689}{AI Act}. \emph{See} \S{}2.1.1 above.}), these GPAI models should also be considered as ``placed on the market'' according to Recital 97. Furthermore, after AI system integration, the provisions described in \S{}2.1 with regard to AI systems would become in scope. These logical steps will be explained in further detail below.

\textbf{Internal deployment often concerns GPAI models}.\footnote{As a purely illustrative example, Google DeepMind recently announced a coding agent, called AlphaEvolve, which integrates Gemini models (\href{https://deepmind.google/blog/alphaevolve-a-gemini-powered-coding-agent-for-designing-advanced-algorithms/}{Google DeepMind}, 2025). Based on information published by the provider, ``AlphaEvolve enhanced the efficiency of Google's data centers, chip design and AI training processes --- including training the large language models underlying AlphaEvolve itself'' (\href{https://deepmind.google/blog/alphaevolve-a-gemini-powered-coding-agent-for-designing-advanced-algorithms/}{Google DeepMind}, 2025).} It is a common intuition amongst AI researchers that the AI models and systems that providers elect to deploy internally are often the most capable at any given time.\footnote{\emph{See} \href{https://arxiv.org/abs/2504.12170}{Stix et al.}, 2025, at 7-9; \href{https://static1.squarespace.com/static/64edf8e7f2b10d716b5ba0e1/t/687e324254b8df665abc5664/1753100867033/Managing+Risks+from+Internal+AI+Systems.pdf}{Acharya and Delaney}, 2025, at 8 (``the best AIs at any given time are often internal \ldots{} may be significantly better than public ones''). \emph{See} footnote 7.} This is particularly true when internal AI models and systems are used for autonomous AI R\&D. This intuition appears to be confirmed in the text of the COP, which considers ``capabilities to automate AI research and development'' in GPAI models as ``sources of systemic risks''\footnote{Appendix 1.3, \href{https://digital-strategy.ec.europa.eu/en/policies/contents-code-gpai}{COP}.} and acknowledges that GPAI models can be ``used'' ``in the development, oversight, and/or evaluation of models.''\footnote{Measure 7.1, \href{https://digital-strategy.ec.europa.eu/en/policies/contents-code-gpai}{COP}. \emph{See} also Recital 110, \href{https://eur-lex.europa.eu/legal-content/EN/TXT/PDF/?uri=OJ:L_202401689}{AI Act}.} In other words, according to the COP, the capabilities necessary to automate AI R\&D can be found in GPAI models. Of course, there are exceptions, as AI providers may as well deploy internally models that are not GPAI models.\footnote{An illustrative example of non-GPAI models being deployed internally could be the use of AlphaFold by Google DeepMind and Isomorphic Labs (\href{https://www.isomorphiclabs.com/articles/a-glimpse-of-the-next-generation-of-alphafold}{Isomorphic Labs}, 2023).}

\textbf{Internal deployment, and specifically autonomous AI R\&D, necessitates or benefits from integration of GPAI models into AI systems}. In order to be maximally useful to its providers for a variety of potential internal uses, including autonomous AI R\&D, a GPAI model does not only need to be sufficiently capable.\footnote{For an overview on the relationship between affordances and capabilities, \emph{see} \href{https://arxiv.org/pdf/2511.15846}{Stix et al}., 2025, at 24. On internal AI models and systems specifically, \emph{see} \emph{also} \href{https://arxiv.org/abs/2504.12170}{Stix et al.}, 2025, at 10-12 (describing read, write, and execute permissions that internal AI models and systems may necessitate with respect to their own hardware, weights, architecture, training, or oversight mechanisms or on those of their successors).} A GPAI model also needs to be sufficiently `enabled' through affordances and permissions to influence the real world.\footnote{\emph{See} footnote 88 above.} This `enablement' occurs by integrating AI models (including GPAI models) into existing or new AI systems. As the International Scientific Report on the Safety of Advanced AI explains, an AI system is ``[a]n integrated setup that combines one or more AI models with other components.''\footnote{\href{https://internationalaisafetyreport.org/sites/default/files/2025-10/international_ai_safety_report_2025_english.pdf}{International Scientific Report on the Safety of Advanced AI}, 2025, at 31.} In other words, AI systems are a ``slight generalization of AI models'' that includes not only the weights and architecture of an AI model, but also a ``broader set of system parameters,'' which strongly influence the capabilities of AI systems.\footnote{\href{https://www.apolloresearch.ai/u/2025/11/A-Causal-Framework-for-AI-Regulation-and-Auditing-.pdf}{Sharkey et al.}, 2023, at 4.} This reflects the definition of ``AI system'' contained in Article 3(1) of the AI Act, according to which AI systems are ``machine-based system[s]'' that ``operate with varying levels of autonomy'' and ``can influence physical or virtual environments.''\footnote{Article 3(1), \href{https://eur-lex.europa.eu/legal-content/EN/TXT/PDF/?uri=OJ:L_202401689}{AI Act}. \emph{See also} Recital 12, \href{https://eur-lex.europa.eu/legal-content/EN/TXT/PDF/?uri=OJ:L_202401689}{AI Act}.} The European Commission's Guidelines on the definition of ``AI system'' further clarify that ``[a]ll AI systems'' ``require'' ``model training'' as well as ``hardware'' and ``software components'' to interact with its external environment, including ``input/output interfaces.''\footnote{\emph{See} Section II.1-2, Commission Guidelines on the definition of an AI system established by Regulation (EU) 2024/1689 (AI Act) (\href{https://digital-strategy.ec.europa.eu/en/library/commission-publishes-guidelines-ai-system-definition-facilitate-first-ai-acts-rules-application}{European Commission}, 2025, at 2-3).} Without system parameters and environmental resources and opportunities for affecting the world that are available to AI systems (often referred to as `affordances'\footnote{\href{https://www.apolloresearch.ai/u/2025/11/A-Causal-Framework-for-AI-Regulation-and-Auditing-.pdf}{Sharkey et al.}, 2023, at 5.}), AI systems could hardly ``influence physical or virtual environments.''\footnote{Article 3(1), \href{https://eur-lex.europa.eu/legal-content/EN/TXT/PDF/?uri=OJ:L_202401689}{AI Act}.} For example, in order to automate AI R\&D, internal AI systems may need access to multiple resources, including access to compute, to energy, to sensors to gather information about the environment, and to a wide range of information (e.g., a provider's code including algorithmic information, as well as a provider's safety, security, and technical oversight mechanisms).\footnote{\href{https://arxiv.org/abs/2504.12170}{Stix et al.}, 2025, at 10-12.}

\textbf{GPAI models are considered as ``placed on the market'' after they are integrated into an AI system (and the AI system is ``put into service'')}. Recital 97 clarifies that ``[w]hen the provider of a general-purpose AI model \emph{integrates an own model into its own AI system} that is made available on the market or put into service,\footnote{\emph{See} \S{}\S{}2.1.1-2.1.2., discussing arguments in favor of considering internal AI systems as put in service (\S{}2.1.1) and placed on the market under Article 2(1)(a) of the AI Act (\S{}2.1.2).} that model \emph{should be considered to be placed on the market} and, therefore, the obligations in this Regulation for models should continue to apply in addition to those for AI systems.''\footnote{Recital 97, \href{https://eur-lex.europa.eu/legal-content/EN/TXT/PDF/?uri=OJ:L_202401689}{AI Act} (emphasis added). \emph{See also} Court of Justice of the European Union, \href{https://eur-lex.europa.eu/resource.html?uri=cellar:bff0427d-b75d-46bf-8015-7eeddf6f9ca8.0002.06/DOC_1&format=PDF}{C-215/88} (clarifying that ``a recital in the preamble to a regulation may cast light on the interpretation to be given to a legal rule''---arguably, Article 3 of the AI Act in this case).} In other words, once a GPAI model is integrated into an AI system\footnote{\emph{See} Recital 100, \href{https://eur-lex.europa.eu/legal-content/EN/TXT/PDF/?uri=OJ:L_202401689}{AI Act} (``When a general-purpose AI model is integrated into or forms part of an AI system, this system should be considered to be a general-purpose AI system when, due to this integration, this system has the capability to serve a variety of purposes'').} and that AI system is put into service, the AI Act considers the GPAI model as ``placed on the market,''\footnote{Article 2(1)(a), \href{https://eur-lex.europa.eu/legal-content/EN/TXT/PDF/?uri=OJ:L_202401689}{AI Act}.} and the GPAI model and relevant AI systems consequently fall within the scope of the AI Act under Article 2(1)(a).

Therefore, under this argument, even if GPAI models were not subject to the AI Act before their placement on the market, these GPAI models would still fall within the remit of the AI Act once they are integrated into an AI system for the purposes of deploying them internally (because that would constitute market placement under the AI Act). In other words, under this argument, \textbf{Article 2(1)(a) covers GPAI models from the moment they are integrated into an AI system, which is put into service}. By contrast, internal GPAI models that are \emph{not} integrated into an internal AI system that is put into service would not be considered as placed on the market and, therefore, would remain outside the scope of the AI Act \emph{before} public release (or AI system integration). Based on the considerations above regarding the correlation between system integration and the opportunity for a GPAI model to influence the real world, the scope of a potential carve-out of internal GPAI models from the scope of the AI Act under Article 2(1)(a) is rather limited.\footnote{More speculatively, this might also raise the question as to whether the EU legislator simply saw no need to include within Article 2(1)(a) the option: `putting into service general-purpose AI models.'} This is particularly true considering that, on average, system integration occurs \emph{before} (and not after) public release.\footnote{\emph{See} the \href{https://internationalaisafetyreport.org/sites/default/files/2025-10/international_ai_safety_report_2025_english.pdf}{International Scientific Report on the Safety of Advanced AI}, 2025, at 30.} In other words, once a GPAI model is placed on the market, chances are it has already been integrated into an AI system.

Third, and finally, if Article 2(1)(a) were interpreted to exclude internally-deployed GPAI models (along with the relevant internal AI systems into which GPAI models may be integrated) from the scope of the AI Act, this interpretation could clash with the content and purpose of, among others, Article 2(1)(a)-(c) and Recital 97, and it could also create an enforcement distortion. As mentioned above, internal AI systems arguably fall within the scope of the AI Act under Article 2(1)(a)-(c).\footnote{\emph{See} \S{}\S{}2.1.1-2.1.3 above.} If that is the case, and if internal GPAI models were considered to be excluded from the scope of the AI Act (along with the relevant internal AI systems into which GPAI models may be integrated), the AI Act: (i) would apply to AI systems embedding AI models that are \emph{not} GPAI models, i.e., AI models that are less capable and potentially less concerning than GPAI models;\footnote{\emph{See} Article 3(63), \href{https://eur-lex.europa.eu/legal-content/EN/TXT/PDF/?uri=OJ:L_202401689}{AI Act} (defining ``general-purpose AI model''); Recital 110, \href{https://eur-lex.europa.eu/legal-content/EN/TXT/PDF/?uri=OJ:L_202401689}{AI Act} (clarifying that ``[g]eneral-purpose AI models could pose systemic risks''); Recital 97, \href{https://eur-lex.europa.eu/legal-content/EN/TXT/PDF/?uri=OJ:L_202401689}{AI Act} (mentioning GPAI models' ``generality and ... capability to competently perform a wide range of distinct tasks'').} and (ii) would not apply to internally-deployed GPAI models and systems, which are generally more capable and thus have a greater potential to pose threats, including systemic risk.\footnote{On the other hand, it should be noted that this outcome might be consistent with the fact that obligations on providers of GPAI models ``can be considered a `light' version of the obligations for AI systems'' (\href{https://law-store.wolterskluwer.com/s/product/the-eu-artificial-intelligence-ai-act-a-commentary/01tPg000007gkK9IAI?srsltid=AfmBOooB8DDUCEJELRg4SkWxkoS853DEceZpZ3x659nmpw3IsLNZ6WuM}{Van Eecke and Regenhardt}, 2024, at 35).}

\subsection*{3. Arguments Opposing or Restricting the Application of the AI Act to Internal Deployment}
\addcontentsline{toc}{subsection}{3. Arguments Opposing or Restricting the Application of the AI Act to Internal Deployment}

\S{}2 above explored potential arguments supporting the inclusion of internal deployment within the scope of the AI Act. This section does the opposite: it explores and stress-tests potential arguments that support the exclusion of internal deployment from the scope of the AI Act, or restrict the application of the AI Act to only some forms of internal uses. Specifically:

\begin{itemize}
  \item \textbf{\S{}3.1} concentrates on the expression ``AI systems or AI models, including their output, specifically developed and put into service for the sole purpose of scientific research and development'' in Article 2(6).
  \item \textbf{\S{}3.2} concentrates on the expression ``any research, testing or development activity regarding AI systems or AI models prior to their being placed on the market or put into service'' in Article 2(8). 
\end{itemize}

As it will be described below, the potential arguments based on these provisions hardly support a totalitarian exclusion of internal deployment from the scope of the AI Act, but rather enable select carve-outs from the scope of the Act.\footnote{\emph{See} \S{}\S{}3.2-3.3 below. \emph{See also} Figure 1 above, at 1.}

\subsection*{3.1 Article 2(6): ``AI systems or AI models, including their output, specifically developed and put into service for the sole purpose of scientific research and development.''}
\addcontentsline{toc}{subsection}{3.1 Article 2(6): ``AI systems or AI models, including their output, specifically developed and put into service for the sole purpose of scientific research and development.''}

A first argument to restrict the inclusion of internal deployment within the scope of the AI Act is based on Article 2(6). This Article arguably introduces the most interesting open question on the application of the AI Act to internal deployment.\footnote{\emph{See}, for instance, \href{https://law-store.wolterskluwer.com/s/product/the-eu-artificial-intelligence-ai-act-a-commentary/01tPg000007gkK9IAI?srsltid=AfmBOooB8DDUCEJELRg4SkWxkoS853DEceZpZ3x659nmpw3IsLNZ6WuM}{Van Eecke and Regenhardt}, 2024, at 44 (``it is difficult to assess when research, testing and development is subjected to the scope of the AI Act'').} Under Article 2(6), the AI Act ``does not apply to AI systems \emph{or} AI models, including their output, \emph{specifically developed and put into service for the sole purpose of scientific research and development}.''\footnote{Article 2(6), \href{https://eur-lex.europa.eu/legal-content/EN/TXT/PDF/?uri=OJ:L_202401689}{AI Act} (emphasis added). \emph{See also} Recital 109, \href{https://eur-lex.europa.eu/legal-content/EN/TXT/PDF/?uri=OJ:L_202401689}{AI Act} (``Compliance with the obligations applicable to the providers of general-purpose AI models should be commensurate and proportionate to the type of model provider, \emph{excluding the need for compliance for persons who develop or use models for \ldots{} scientific research purposes};'' emphasis added).}

This Article seemingly introduces an exception to the general applicability of the AI Act to internal deployment. In other words, if Article 2(1)(a)-(c) were interpreted as described in \S{}\S{}2.1-3.1 above, Article 2(6) could be used to argue that, while internal deployment is generally covered by the AI Act, AI models and systems that are ``specifically'' developed and deployed for ``scientific research and development'' are exempted. This section addresses this exception and examines its scope.

Two expressions appear particularly relevant when interpreting Article 2(6): (i) ``specifically developed and put into service;'' and (ii) ``sole purpose of scientific research and development.''\footnote{Article 2(6), \href{https://eur-lex.europa.eu/legal-content/EN/TXT/PDF/?uri=OJ:L_202401689}{AI Act}.} First, ``\textbf{specifically developed and put into service}'' is not defined by the AI Act. The word ``specifically'' and the conjunction ``or'' (``AI systems \emph{or} AI models'')\footnote{Under Article 2(6), the AI Act ``does not apply to AI systems \emph{or} AI models, including their output, specifically developed and put into service for the sole purpose of scientific research and development'' (emphasis added on the ``or'' conjunction). \emph{See} Article 2(6), \href{https://eur-lex.europa.eu/legal-content/EN/TXT/PDF/?uri=OJ:L_202401689}{AI Act}.} appear to suggest that Article 2(6) concerns only AI models or AI systems that are \textbf{developed \emph{ad hoc} to serve one exclusive purpose} (i.e., the ``sole purpose of scientific research and development'') \textbf{\emph{and}} \textbf{deployed exclusively for that purpose}.\footnote{Recital 109 of the AI Act excludes ``the need for compliance'' ``with the obligations applicable to the providers of general-purpose AI models'' for ``persons who develop \emph{or} use models for \emph{\ldots{}} scientific research purposes'' (Recital 109, \href{https://eur-lex.europa.eu/legal-content/EN/TXT/PDF/?uri=OJ:L_202401689}{AI Act}; emphasis added). Through the `or' conjunction, Recital 109 seems to suggest that it would be sufficient for a GPAI model to be \emph{deployed} for the sole purpose of scientific R\&D in order to fall outside the scope of the AI Act. In other words, for the purposes of Recital 109, a GPAI model would not need to be \emph{developed and deployed} for scientific R\&D, but, for instance, it would only need to be \emph{deployed} for that purpose. This interpretation appears in conflict with EU law, as Recital 109 would derogate from the actual provisions of the act in question (i.e., Article 2(6), \href{https://eur-lex.europa.eu/legal-content/EN/TXT/PDF/?uri=OJ:L_202401689}{AI Act}). \emph{See}, for instance, Court of Justice of the European Union, \href{https://curia.europa.eu/juris/document/document.jsf?text=\%2522legal\%2Bforce\%2522\%2Brecital&docid=295079&pageIndex=0&doclang=EN&mode=req&dir=&occ=first&part=1&cid=669610\#ctx1}{C-423/23 2025}, \href{https://curia.europa.eu/juris/document/document.jsf?text=\%2522legal\%2Bforce\%2522\%2Brecital&docid=293845&pageIndex=0&doclang=EN&mode=req&dir=&occ=first&part=1&cid=669610\#ctx1}{C-664/23 2024}, \href{https://curia.europa.eu/juris/document/document.jsf?text=&docid=234325&pageIndex=0&doclang=EN&mode=lst&dir=&occ=first&part=1&cid=670078}{C-302/19 2020}, \href{https://curia.europa.eu/juris/document/document.jsf?text=&docid=221805&pageIndex=0&doclang=EN&mode=lst&dir=&occ=first&part=1&cid=670435}{C-418/18 2019}, \href{https://curia.europa.eu/juris/document/document.jsf?text=&docid=153817&pageIndex=0&doclang=en&mode=lst&dir=&occ=first&part=1&cid=673080}{C-345/13 2014}, \href{https://curia.europa.eu/juris/showPdf.jsf?text=&docid=56151&pageIndex=0&doclang=EN&mode=lst&dir=&occ=first&part=1&cid=671551}{C-136/04 2005}, \href{https://curia.europa.eu/juris/showPdf.jsf?text=&docid=44220&pageIndex=0&doclang=en&mode=lst&dir=&occ=first&part=1&cid=16849126}{C-162/97 1998}.}

Second, ``\textbf{sole purpose of scientific research and development}'' is also not defined by the AI Act. ``Sole'' contributes to the idea that AI models and systems that fall into the exception set by Article 2(6) must have only one purpose. In other words, ``sole'' excludes from the spectrum of Article 2(6) AI models and systems that are deployed for multiple purposes.\footnote{\emph{See}, for instance, \href{https://arxiv.org/pdf/2302.02337}{Hacker et al}., 2023, at 6 (``this research exemption arguably does not apply anymore once the system is released into the wild, as any public release likely does not have scientific research and development as its `sole purpose''').} ``\textbf{\emph{Scientific}} \textbf{research and development}'' \textbf{must therefore be the only purpose of training and deploying}. This is further confirmed by Recital 25, which clarifies that ``without prejudice to the exclusion of AI systems specifically developed and put into service for the sole purpose of scientific research and development, \emph{any other AI system that may be used for the conduct of any research and development} activity should remain subject to the provisions of this Regulation.'' In other words, not all the AI models and systems deployed for research and development are exempt---only the ones specifically developed and trained for ``\emph{scientific}'' research and development.\footnote{On this point, \emph{see} \href{https://academic.oup.com/grurint/article/74/10/903/8238058?login=false}{Finck}, 2025.}

The considerations above allow us to reach an intermediate conclusion: Article 2(6) of the AI Act does not exempt internal deployment from the scope of the AI Act. If anything, Article 2(6) creates a narrow exception for internal ``AI systems or AI models'' that a provider purposefully develops and deploys for one exclusive purpose: scientific research and development.\footnote{Notably, Article 2(6) uses the expression ``AI systems \emph{or} AI models'' (emphasis added). This creates an open question as to how the AI Act should be interpreted with regard to situations in which: (i) an AI \emph{model} is not developed \emph{ad hoc} for scientific research and development (e.g., a GPAI model); however, (ii) the AI \emph{system} in which the model is integrated is developed \emph{ad hoc} for scientific research and development. In this respect, the conjunction `or' could suggest that if \emph{either} one between the AI model or the AI system is ``specifically developed and put into service for the sole purpose of scientific research and development'' (Article 2(6), \href{https://eur-lex.europa.eu/legal-content/EN/TXT/PDF/?uri=OJ:L_202401689}{AI Act}), then that AI model or that AI system could fall outside of the AI Act's scope.} The main challenge of Article 2(6), therefore, becomes understanding what ``\emph{scientific} research and development'' refers to in the context of AI,\footnote{Article 2(6), \href{https://eur-lex.europa.eu/legal-content/EN/TXT/PDF/?uri=OJ:L_202401689}{AI Act} (emphasis added). Importantly, it is also possible that ``scientific research and development'' may not refer to AI R\&D at all, but rather and more broadly to projects of scientific nature.} and specifically whether it can be relied on to exclude from the AI Act's scope AI models and systems that are specifically trained and deployed to automate AI R\&D.

Importantly, while not defining ``\emph{scientific} research and development,'' the AI Act appears to \textbf{juxtapose} ``\textbf{\emph{scientific}} \textbf{research and development}''\footnote{Article 2(6), \href{https://eur-lex.europa.eu/legal-content/EN/TXT/PDF/?uri=OJ:L_202401689}{AI Act} (emphasis added).} \textbf{to} ``\textbf{\emph{product-oriented}} \textbf{research}.''\footnote{Recital 25, \href{https://eur-lex.europa.eu/legal-content/EN/TXT/PDF/?uri=OJ:L_202401689}{AI Act}.} In particular, Recital 25 provides that, ``[a]s regards \emph{product-oriented research}, testing and development activity regarding AI systems or models, the provisions of this Regulation should also not apply \emph{prior to those systems and models being put into service or placed on the market}.''\footnote{Recital 25, \href{https://eur-lex.europa.eu/legal-content/EN/TXT/PDF/?uri=OJ:L_202401689}{AI Act} (emphasis added).} In doing so, Recital 25 clarifies that product-oriented research is in fact covered by the AI Act and is subject only to the exception set forth under Article 2(8) (\emph{see} \S{}3.2 below). Therefore, as will be explained below, \emph{after} an AI model or system for product-oriented research is put into service (i.e., deployed internally or externally), it is subject to the obligations set out in the AI Act. To summarize, it appears that:

\begin{itemize}
  \item The AI Act differentiates between \emph{scientific} R\&D and \emph{product}-oriented R\&D (Article 2(6) and Recital 25 of the AI Act).
\end{itemize}

\begin{itemize}
  \item Only AI models or systems specifically developed and deployed internally with the sole, exclusive purpose of doing \emph{scientific} R\&D are exempted under Article 2(6) of the AI Act.
\end{itemize}

\begin{itemize}
  \item By contrast, AI models and systems developed and deployed internally with the purpose of doing \emph{product-oriented} R\&D are not exempted (subject to Article 2(8) and Recital 25; \emph{see} \S{}3.2 below). 
\end{itemize}

As mentioned above, the AI Act does not define ``scientific research and development.''\footnote{Article 2(6), \href{https://eur-lex.europa.eu/legal-content/EN/TXT/PDF/?uri=OJ:L_202401689}{AI Act}.} The AI Act does not define ``product-oriented research and development'' either.\footnote{Recital 25, \href{https://eur-lex.europa.eu/legal-content/EN/TXT/PDF/?uri=OJ:L_202401689}{AI Act}.} To shed light on the difference between these two kinds of research and development, a helpful point of reference could be Regulation 1907/2006 of December 18, 2006 concerning the Registration, Evaluation, Authorisation and Restriction of Chemicals (REACH).\footnote{Other authors have offered parallels between the AI Act and the EU General Data Protection Regulation (GDPR; \href{https://eur-lex.europa.eu/eli/reg/2016/679/oj}{Regulation (EU) 2016/679 of April 27, 2016}). \emph{See} \href{https://academic.oup.com/grurint/article/74/10/903/8238058?login=false}{Finck}, 2025. Specifically, Recital 159 of the GDPR clarifies that ``the processing of personal data for scientific research purposes should be interpreted \emph{in a broad manner} including for example \emph{technological development and demonstration}, \emph{fundamental research}, \emph{applied research} and privately funded research'' (\href{https://eur-lex.europa.eu/eli/reg/2016/679/oj}{Regulation (EU) 2016/679 of April 27, 2016}; emphasis added).} Like the AI Act, the REACH Regulation similarly distinguishes between Product and Process Orientated Research and Development (`PPORD') and Scientific Research and Development (SR\&D), and exempts the latter from some of the obligations and restrictions set forth by the Regulation.\footnote{Article 3(22)-(23), and Article 67(1), Regulation (EC) 1907/2006 of December 18, 2006 (\href{https://eur-lex.europa.eu/legal-content/EN/TXT/PDF/?uri=CELEX:32006R1907}{REACH Regulation}).} Essentially, just like the AI Act, the REACH Regulation has a dual approach to product-oriented research (which it calls PPORD) and scientific research (which it calls SR\&D). 

Specifically, under the REACH Regulation, \textbf{Product and Process Orientated Research and Development} (i.e., PPORD) is defined as ``any scientific development \emph{related to product development} or the further development of a substance.''\footnote{Article 3(22), \href{https://eur-lex.europa.eu/legal-content/EN/TXT/PDF/?uri=CELEX:32006R1907}{REACH Regulation}.} Relevant guidance by the European Chemical Agency (ECHA) clarifies that product-oriented research includes a wide range of activities aimed at developing or proving the feasibility of new products and processes, and improving production efficiency. ECHA's guidance provides the following examples of product-oriented research: ``campaign(s) for the scaling-up or improvement of a production process in a pilot plant or in the full-scale production, or the investigation of the fields of applications for that substance.''\footnote{European Chemicals Agency (ECHA), Guidance on Scientific Research and Development (SR\&D) and Product and Process Orientated Research and Development (PPORD) (\href{https://echa.europa.eu/documents/10162/2324906/ppord_en.pdf/22a12900-ad27-454c-aedd-82972ef2f675}{ECHA Guidance}, 2017), at 7-8 (emphasis added).} 

\textbf{Scientific Research and Development} (i.e., SR\&D), on the other hand, is defined as ``any scientific experimentation, analysis or \ldots{} research carried out \textbf{\emph{under controlled conditions}}'' within a certain volume (``less than 1 tonne per year'').\footnote{Article 3(23), \href{https://eur-lex.europa.eu/legal-content/EN/TXT/PDF/?uri=CELEX:32006R1907}{REACH Regulation}\emph{. See also} Recital 28, \href{https://eur-lex.europa.eu/legal-content/EN/TXT/PDF/?uri=CELEX:32006R1907}{REACH Regulation} (``scientific research and development normally takes place in quantities below 1 tonne per year'').} ECHA's guidance clarifies that scientific research and development may include ``experimental research or analytical activities \textbf{\emph{at a laboratory scale}}'' ``as well as the use of the substance in monitoring and routine quality control or \textbf{\emph{in vitro diagnostics}} at a laboratory scale under controlled conditions.''\footnote{\href{https://echa.europa.eu/documents/10162/2324906/ppord_en.pdf/22a12900-ad27-454c-aedd-82972ef2f675}{ECHA Guidance}, 2017, at 7.} ECHA's guidance also clarifies that research and development activity occurs ``under controlled conditions'' if ``procedures and measures are in place to minimise or control exposure and potential risks from exposure of humans and the environment to the substance.''\footnote{\href{https://echa.europa.eu/documents/10162/2324906/ppord_en.pdf/22a12900-ad27-454c-aedd-82972ef2f675}{ECHA Guidance}, 2017, at 8 (``This may include, for example, limitation of uses to qualified persons having access to the substance, or collection and disposal of waste'').} 

In other words, the REACH Regulation appears to differentiate Product and Process Orientated Research and Development and Scientific Research and Development (SR\&D) through a combination of two main parameters.\footnote{The REACH Regulation does not appear to clarify if any of these two factors should be prioritized.} First, there is the \textbf{scale} of the R\&D activity. Scientific Research and Development involves quantities ``less than 1 tonne per year,'' ``under controlled conditions,'' and ``at a laboratory scale'' (such as ``in vitro diagnostics''),\footnote{Article 3(23) and Recital 28, \href{https://eur-lex.europa.eu/legal-content/EN/TXT/PDF/?uri=CELEX:32006R1907}{REACH Regulation}; \href{https://echa.europa.eu/documents/10162/2324906/ppord_en.pdf/22a12900-ad27-454c-aedd-82972ef2f675}{ECHA Guidance}, 2017, at 8.} whereas Product and Process Orientated Research and Development can be either ``in a pilot plant'' or in ``full-scale production.''\footnote{\href{https://echa.europa.eu/documents/10162/2324906/ppord_en.pdf/22a12900-ad27-454c-aedd-82972ef2f675}{ECHA Guidance}, 2017, at 7 (emphasis added).} Intuitions around the Scientific Research and Development activity could be easily translated to scientific AI R\&D: this type of AI R\&D should be undertaken in controlled conditions and at laboratory scale (for example, in vitro).\footnote{Arguably, the fact that scientific R\&D occurs in `controlled conditions' may be one of the reasons why this type of R\&D is less regulated than product-oriented R\&D. If a scientific experiment is undertaken in properly controlled conditions, the negative impact of potential incidents would plausibly be more contained than if conditions were not controlled.} 

Second, there is the \textbf{purpose} of the R\&D activity. Product and Process Orientated Research and Development aims at ``product development or the further development of a substance,''\footnote{Article 3(22), \href{https://eur-lex.europa.eu/legal-content/EN/TXT/PDF/?uri=CELEX:32006R1907}{REACH Regulation}.} whereas Scientific Research and Development aims at ``experimentation, analysis,''\footnote{Article 3(23), \href{https://eur-lex.europa.eu/legal-content/EN/TXT/PDF/?uri=CELEX:32006R1907}{REACH Regulation}\emph{. See also} Recital 28, \href{https://eur-lex.europa.eu/legal-content/EN/TXT/PDF/?uri=CELEX:32006R1907}{REACH Regulation} (``scientific research and development normally takes place in quantities below 1 tonne per year'').} ``monitoring and \ldots{} diagnostics.''\footnote{\href{https://echa.europa.eu/documents/10162/2324906/ppord_en.pdf/22a12900-ad27-454c-aedd-82972ef2f675}{ECHA Guidance}, 2017, at 7.} Translating these concepts to AI R\&D, product-oriented research and development could be interpreted to include any scientific development related to the enhancement and scaling of AI capabilities, AI products, and/or the training pipeline;\footnote{An illustrative example of scientific activity aimed at the enhancement and scaling of AI capabilities, AI products, and/or the training pipeline could be the discovery of Reinforcement Learning (RL) pipelines that patch existing vulnerabilities in current AI systems, or solutions to improve Graphics Processing Unit (GPU) utilization.} by contrast, scientific R\&D could be AI R\&D that is not product-oriented. 

These two parameters (i.e., the scale and the purpose of the R\&D activity) could help orient the interpretation of ``scientific research and development'' in Article 2(6) of the AI Act, offering an initial point of reference for determining what research and development activities are \emph{scientific} R\&D and what are instead \emph{product-oriented} R\&D. For instance, by applying these two parameters, it would be possible to observe that the parameter of \textbf{scale} for scientific research and development as extrapolated from the REACH Regulation (i.e., ``under controlled conditions,'' and ``at a laboratory scale'' such as ``in vitro''\footnote{Article 3(23) and Recital 28, \href{https://eur-lex.europa.eu/legal-content/EN/TXT/PDF/?uri=CELEX:32006R1907}{REACH Regulation}; \href{https://echa.europa.eu/documents/10162/2324906/ppord_en.pdf/22a12900-ad27-454c-aedd-82972ef2f675}{ECHA Guidance}, 2017, at 8.}) would be met, for instance, by what AI researchers define as `model organisms.'\footnote{\emph{See} \href{https://www.alignmentforum.org/posts/ChDH335ckdvpxXaXX/model-organisms-of-misalignment-the-case-for-a-new-pillar-of-1}{Hubinger et al.,} 2023; \href{https://arxiv.org/pdf/2401.05566}{Hubinger et al}., 2024, at 8-9.} Model organisms are, literally, ``\emph{in vitro} demonstrations of the kinds of failures that might pose existential threats.''\footnote{\href{https://www.alignmentforum.org/posts/ChDH335ckdvpxXaXX/model-organisms-of-misalignment-the-case-for-a-new-pillar-of-1}{Hubinger et al.,} 2023. \emph{See also} \href{https://arxiv.org/pdf/2401.05566}{Hubinger et al}., 2024.} Another tentative example of research and development activities ``under controlled conditions'' and ``at a laboratory scale'' is the training of AI models or systems for mechanistic interpretability research.\footnote{Mechanistic interpretability is a field of research focused on reverse engineer neural networks. \emph{See generally} \href{https://arxiv.org/pdf/2501.16496}{Sharkey et al}., 2024.} 

Furthermore, these two illustrative examples (i.e., model organisms and mechanistic interpretability) also fit the parameter of \textbf{purpose} as extrapolated from the REACH Regulation. Both model organisms and mechanistic interpretability are usually detached from (or, at least, not directly aimed at) the enhancement and scaling of AI capabilities, AI products, and/or the training pipeline. Model organisms are used to offer scientific proof-of-concept, within controlled environments, of the existence or likelihood of specific behaviors in AI models and systems.\footnote{\emph{See} footnotes 173-176 below.} Recent examples include AI researchers: training `backdoored' model organisms, and then applying safety training, to evaluate whether the backdoor behavior persists in these model organisms;\footnote{\emph{See} \href{https://arxiv.org/pdf/2401.05566}{Hubinger et al}., 2024.} training model organisms to study the effects on misalignment of fine-tuning AI models on narrowly harmful datasets;\footnote{\emph{See} \href{https://arxiv.org/pdf/2502.17424}{Betley} \href{https://arxiv.org/pdf/2502.17424}{et al}., 2025; \href{https://arxiv.org/pdf/2506.11613}{Turner} \href{https://www.arxiv.org/pdf/2506.19823}{et al}., 2025; \href{https://www.arxiv.org/pdf/2506.19823}{Wang et al}., 2025.} training an ``evaluation-aware'' model organism to validate the effect of activation steering on evaluation awareness;\footnote{\emph{See} \href{https://arxiv.org/pdf/2510.20487}{Hua et al}., 2025.} or training an AI model with a hidden objective to then undertake alignment audits.\footnote{\emph{See} \href{https://arxiv.org/pdf/2503.10965}{Marks et al}., 2025.} Mechanistic interpretability research aims to reverse engineer a model's computations and is often not immediately useful for product development.\footnote{\emph{See} \href{https://arxiv.org/pdf/2511.13653}{Gao et al}., 2025; \href{https://openai.com/index/understanding-neural-networks-through-sparse-circuits/}{OpenAI}, 2025.} For instance, OpenAI recently trained AI models with an architecture similar to (now largely outdated) GPT-2, and forced most weights to be zero, in order to ``substantially disentangle[] the model's internal computations.''\footnote{\emph{See} \href{https://arxiv.org/pdf/2511.13653}{Gao et al}., 2025; \href{https://openai.com/index/understanding-neural-networks-through-sparse-circuits/}{OpenAI}, 2025. Another recent example is \href{https://arxiv.org/abs/2505.13787}{Cundy and Gleave}, 2025, in which researchers trained a model ``incorporating a lie detector into the labelling step of LLM post-training'' in order to ``evaluate whether the learned policy is genuinely more honest, or instead learns to fool the lie detector while remaining deceptive.''} An additional example that can help shed light on an AI model used purely for scientific experimentation is the AI model that Google DeepMind developed in collaboration with Howard Hughes Medical Institute (HHMI) to simulate how a fruit fly walks, flies, and behaves to understand how brain, body, and environment drive specific behaviors in animals.\footnote{\emph{See} \href{https://www.nature.com/articles/s41586-025-09029-4?utm_source=linkedin&utm_medium=social&utm_campaign=&utm_content=}{Vaxenburg et al}., 2025.} 

Despite being a helpful point of reference, the precise extent to which the definitions in the REACH Regulation and the two parameters of \emph{scale} and \emph{purpose} should inform the interpretation of scientific research and development and product-oriented research and development in the AI Act remains an open question. For instance, it is debatable if and to what extent this interpretation of \emph{scale} could apply to scientific R\&D enabled by highly advanced AI systems or AI models, considering that: (i) there are multiple examples of non-product-oriented scientific research and development projects that are not only large, but arguably on a truly grand-scale (for instance, the Large Hadron Collider at the CERN,\footnote{\emph{See} \href{https://home.web.cern.ch/science/accelerators/large-hadron-collider}{CERN}. The Atlas Collaboration authoring the paper that announced the discovery of the Higgs boson comprised around 3,000 authors (\emph{see} \href{https://arxiv.org/pdf/1207.7214}{ATLAS Collaboration}, 2012; \href{https://atlas.cern/authors/atlas-collaboration}{ATLAS experiment}).} or the Human Brain Project\footnote{\emph{See} \href{https://www.humanbrainproject.eu/en/follow-hbp/news/2023/09/28/human-brain-project-ends-what-has-been-achieved/}{Human Brain Project}, 2023.}); and that (ii) the most advanced future AI systems or AI models could arguably have the potential to support these grand-scale projects and unlock some of the most complex existing scientific bottlenecks.\footnote{In this respect, \emph{see}, for instance, Section 1 of President Trump's \href{https://www.whitehouse.gov/presidential-actions/2025/11/launching-the-genesis-mission/}{Executive Order 14363 of November 24, 2025}, launching the ``Genesis Mission'' in the United States ``as a dedicated, coordinated national effort to unleash a new age of AI-accelerated innovation and discovery that can solve the most challenging problems of this century.''} Indeed, there are already some early signs that advanced AI models and systems could significantly accelerate scientific discovery,\footnote{\emph{See}, for instance, proteins' complex structures prediction enabled by Google DeepMind's AlphaFold (\href{https://www.nature.com/articles/s41586-021-03819-2}{Jumper et al}., 2021), which led its developers to be awarded a Nobel prize (\href{https://www.nobelprize.org/prizes/chemistry/2024/press-release/}{The Nobel Prize}, 2024). More recently, \emph{see} \href{https://cdn.openai.com/pdf/4a25f921-e4e0-479a-9b38-5367b47e8fd0/early-science-acceleration-experiments-with-gpt-5.pdf}{Bubeck et al}., 2025, especially at 53 and following, for early science acceleration experiments enabled by OpenAI's GPT-5. \emph{See also} Sakana AI's `AI Scientist,' an early proof of concept for fully automatic scientific discovery in machine learning (\href{https://arxiv.org/pdf/2408.06292}{Lu et al}., 2024).} which the European Union may want to harness to accomplish the objective under Article 179(1) of the Treaty on the Functioning of the EU (TFEU).\footnote{Article 179(1), \href{https://eur-lex.europa.eu/legal-content/EN/TXT/HTML/?uri=CELEX:12016E179}{TFEU} (``The Union shall have the objective of strengthening its scientific and technological bases by achieving a European research area in which researchers, scientific knowledge and technology circulate freely, and encouraging it to become more competitive, including in its industry, while promoting all the research activities deemed necessary by virtue of other Chapters of the Treaties.'').} Furthermore, it should be noted that other EU legal frameworks (for instance, the GDPR) do not refer to scale when describing scientific R\&D.\footnote{\emph{See} Recital 159, EU General Data Protection Regulation (GDPR; \href{https://eur-lex.europa.eu/eli/reg/2016/679/oj}{Regulation (EU) 2016/679 of April 27, 2016}). \emph{See also} footnote 153 above.} 

On the other hand, relying exclusively on the \emph{purpose} of the R\&D activity may ultimately render the distinction between scientific and product-oriented R\&D hazy and unclear. For instance, it is unclear whether and to what extent current and future research delivering generalizable knowledge and methods will ultimately contribute to the training of future generations of AI models and systems.\footnote{Consider, for instance, an AI model or system autonomously theorizing and testing a new learning paradigm (e.g., the new `Nested Learning' by \href{https://abehrouz.github.io/files/NL.pdf}{Behrouz et al.}, 2025), which an AI provider then adopts and adapts to be the new architectural backbone of future generations of AI systems and models. \emph{See also} \href{https://law-store.wolterskluwer.com/s/product/the-eu-artificial-intelligence-ai-act-a-commentary/01tPg000007gkK9IAI?srsltid=AfmBOooB8DDUCEJELRg4SkWxkoS853DEceZpZ3x659nmpw3IsLNZ6WuM}{Van Eecke and Regenhardt}, 2024, at 43 (``a research project might have both scientific and practical applications, calling into doubt the applicability of the exemption'').} 

In conclusion, Article 2(6) could be interpreted to refer exclusively to AI models or AI systems that are trained and made available internally within an AI provider's organization to carry out only ``\emph{scientific} research and development.'' To understand the boundaries of scientific research and development, one could compare scientific research and development against product-oriented research and development, mentioned in Recital 25. Read in conjunction with Recital 25, Article 2(6) appears to single out \textbf{internal AI models or systems custom-built and deployed to do research and development that is \emph{not} product-oriented as the sole exception under this Article}. While the boundaries between product-oriented and scientific research and development remain blurry, it is possible to utilize the definitions of Product and Process Orientated Research and Development (PPORD) and Scientific Research and Development (SR\&D) from the REACH Regulation as an initial point of reference. Examining these definitions  suggests that \emph{scale} and \emph{purpose} are important parameters when distinguishing what qualifies as scientific R\&D activity. The REACH Regulation definitions also help us extrapolate some illustrative examples of \emph{scientific} (e.g., model organisms, mechanistic interpretability) and \emph{product-oriented} (e.g., RL pipelines to patch vulnerabilities in current AI systems, or solutions to improve GPU utilization) AI R\&D.

\subsection*{3.2 Article 2(8): ``Any research, testing or development activity regarding AI systems or AI models prior to their being placed on the market or put into service.''}
\addcontentsline{toc}{subsection}{3.2 Article 2(8): ``Any research, testing or development activity regarding AI systems or AI models prior to their being placed on the market or put into service.''}

A second argument to oppose or restrict the inclusion of internal deployment is based on Article 2(8) of the AI Act. Under Article 2(8), the AI Act ``does not apply to any research, testing or development activity regarding AI systems or AI models \emph{prior to} their being placed on the market or \emph{put into service}.''\footnote{Article 2(8), \href{https://eur-lex.europa.eu/legal-content/EN/TXT/PDF/?uri=OJ:L_202401689}{AI Act}. \emph{See also} Recital 25, \href{https://eur-lex.europa.eu/legal-content/EN/TXT/PDF/?uri=OJ:L_202401689}{AI Act} (``... it is necessary to ensure that this Regulation does not otherwise affect scientific research and development activity on AI systems or models prior to being placed on the market or put into service'').} This section examines the scope of potential exceptions to the application of the AI Act to internal deployment based on this Article.

A clarification is important on Article 2(8). By contrast with Article 2(6),\footnote{\emph{See} \S{}3.2 above.} Article 2(8) \emph{does not} concern the use of an AI model or system to generate research and development outputs. Instead, Article 2(8) concerns the ``research, testing and development'' activity that \emph{precedes} ``their being placed on the market or put into service'' (i.e., their internal deployment or public release).\footnote{Article 2(8), \href{https://eur-lex.europa.eu/legal-content/EN/TXT/PDF/?uri=OJ:L_202401689}{AI Act}.} In other words, Article 2(6) concerns a \textbf{provider's research and development activities \emph{before} an AI model or system is put into service} (Article 2(1)(a)).\footnote{\emph{See} \S{}2.1.1 above.}

This means that, rather than being an exception to Article 2(1)(a), Article 2(8) is in fact fully consistent with the ``scope'' of the AI Act as delineated in Article 2(1).\footnote{\emph{See} \S{}2.1.1 above.} Under Article 2(1),  the AI Act comes into play once a provider puts into service (i.e., deploys) an AI system. \S{}2.1.1 of this memorandum explored how Article 2(1)(a) could be interpreted to include internal deployment of AI systems. \textbf{Article 2(8) concerns the time before any type of deployment, either internal or external}. This is consistent with the text of Recital 25, according to which ``[t]hat exclusion [i.e., the exception under Article 2(8)] is without prejudice to the obligation to comply with this Regulation where an AI system falling into the scope of this Regulation is placed on the market or \emph{put into service as a result of such research and development activity.}''\footnote{Recital 25, \href{https://eur-lex.europa.eu/legal-content/EN/TXT/PDF/?uri=OJ:L_202401689}{AI Act} (``As regards product-oriented research, testing and development activity regarding AI systems or models, the provisions of this Regulation should also not apply prior to those systems and models being put into service or placed on the market. That exclusion is without prejudice to the obligation to \emph{comply with this Regulation} where an AI system falling into the scope of this Regulation is placed on the market or \emph{put into service as a result of such research and development activity} \ldots{}'') (emphasis added).} In other words, Recital 25 clarifies that, while research and development activity before (internal and external) deployment falls outside the scope of the AI Act, the deployment of the resulting AI system will trigger the application of the Act under Article 2(1)(a).\footnote{\S{}2.1.1 above offers arguments as to why \emph{internal} deployment could trigger the application of the AI Act.}

It remains unclear which precise pre-deployment ``research, testing or development'' activities are  exempted from the scope of the AI Act under Article 2(8).\footnote{Article 2(8), \href{https://eur-lex.europa.eu/legal-content/EN/TXT/PDF/?uri=OJ:L_202401689}{AI Act}.} Recital 25 also broadly refers to ``scientific research and development activity \ldots{} prior to being placed on the market or put into service.''\footnote{Recital 25, \href{https://eur-lex.europa.eu/legal-content/EN/TXT/PDF/?uri=OJ:L_202401689}{AI Act}.} Before an AI model or system is ready for internal and/or external deployment, the typical AI development cycle includes: (i) experimentation and planning (for instance, around model architectures, training algorithms, datasets); (ii) pre-training to develop a base model; (iii) post-training techniques to improve the base model (for instance, fine-tuning, reinforcement learning, and safety training).\footnote{\href{https://internationalaisafetyreport.org/sites/default/files/2025-10/international_ai_safety_report_2025_english.pdf}{International Scientific Report on the Safety of Advanced AI}, 2025, at 30, 33, 34, 41-43.} Depending on the interpretation of Article 2(8), the exemption could include all, or only some, of these activities.

\subsection*{4.  Conclusion}
\addcontentsline{toc}{section}{4.  Conclusion}

This memorandum aims to clarify the interpretation of the AI Act and the extent to which it could apply to internal deployment. In this spirit, while acknowledging the jurisdictional limits of the AI Act, this memorandum put forward several interpretative pathways supporting the inclusion of internal deployment within the scope of the AI Act (\S{}2), and also examined potential objections and exceptions to such inclusion (\S{}3).

Specifically, this memorandum explored and stress-tested the following four arguments in favor of including the internal deployment of AI systems within the scope of the AI Act (\S{}\S{}2.1.1-2.1.4).  \emph{First}, Article 2(1)(a) (``providers \ldots{} putting into service AI systems'') could be interpreted to include within the scope of the AI Act EU-based or foreign AI developers that make an internal AI system available for their own use in the EU (\emph{see} \S{}2.1.1 above). \emph{Second}, Article 2(1)(a) (``providers placing on the market \ldots{} AI systems'') could be interpreted to include within the scope of the AI Act EU-based or foreign AI developers that make an internal AI system available for use on the EU market in a business-related context, including for use in the business of the developer itself (\emph{see} \S{}2.1.2 above). \emph{Third}, Article 2(1)(b) (``deployers of AI systems that have their place of establishment or are located within the Union'') could be interpreted to include within the scope of the AI Act EU-based AI developers or EU-based staff of a foreign developer that make use of an internal AI system (\emph{see} \S{}2.1.3 above). \emph{Fourth}, Article 2(1)(c) (``output produced by the AI system is used in the Union'') could be interpreted to include foreign AI developers that make available in the EU (e.g., to EU-based staff) the output produced by an internal AI system deployed abroad (\emph{see} \S{}2.1.4 above). This memorandum also explored how the AI Act could cover the internal deployment of GPAI models, either through these models' integration into AI systems under Article 2(1) and Recital 97, or through an extensive interpretation of the expression ``placing on the market'' in Article 2(1)(a) (\emph{see} \S{}2.2 and \S{}2.1.1 above).

The application of the AI Act to internally deployed AI models and systems, however, remains an open question. After exploring potential arguments in support of including internal AI models and systems within the remit of the AI Act, this memorandum analyzed arguments against their inclusion as well as potential exceptions or carve-outs (\S{}\S{}3.1-3.2). Specifically, this memorandum examined the following two arguments. \emph{First}, Article 2(6) (``AI systems or AI models, including their output, specifically developed and put into service for the sole purpose of scientific research and development'') could be interpreted to exempt from the scope of the AI Act internal AI models or systems that an AI developer trains and makes available internally for the exclusive purpose of carrying out scientific research and development (i.e., research and development that is not product-oriented and that is undertaken in controlled environments, e.g., `in vitro,' for monitoring and diagnostics) (\emph{see} \S{}3.1 above). \emph{Second}, Article 2(8) (``[a]ny research, testing or development activity regarding AI systems or AI models prior to their being placed on the market or put into service'') could be interpreted to exempt an AI developer's research and development activities \emph{before} an AI model or system is put into service, either internally or externally.

If looked at in isolation, all these provisions of the AI Act may appear as conflicting, and even disjointed. That friction, however, is only superficial. Observing all the examined provisions contextually and systematically, \textbf{the joint interpretation of Articles 2(1), 2(6), 2(8), 3(1), 3(3), 3(4), 3(9), 3(10), 3(11), 3(12), 3(63), and Recitals 12, 13, 21, 25, 97, 109, and 110 can not only be brought to coherence, but these provisions can in fact be viewed as complementary to each other}.

Figure 1 below visually represents how these provisions complement and intersect with each other.

\begin{figure}[H]
\centering
\includegraphics[width=0.8\textwidth]{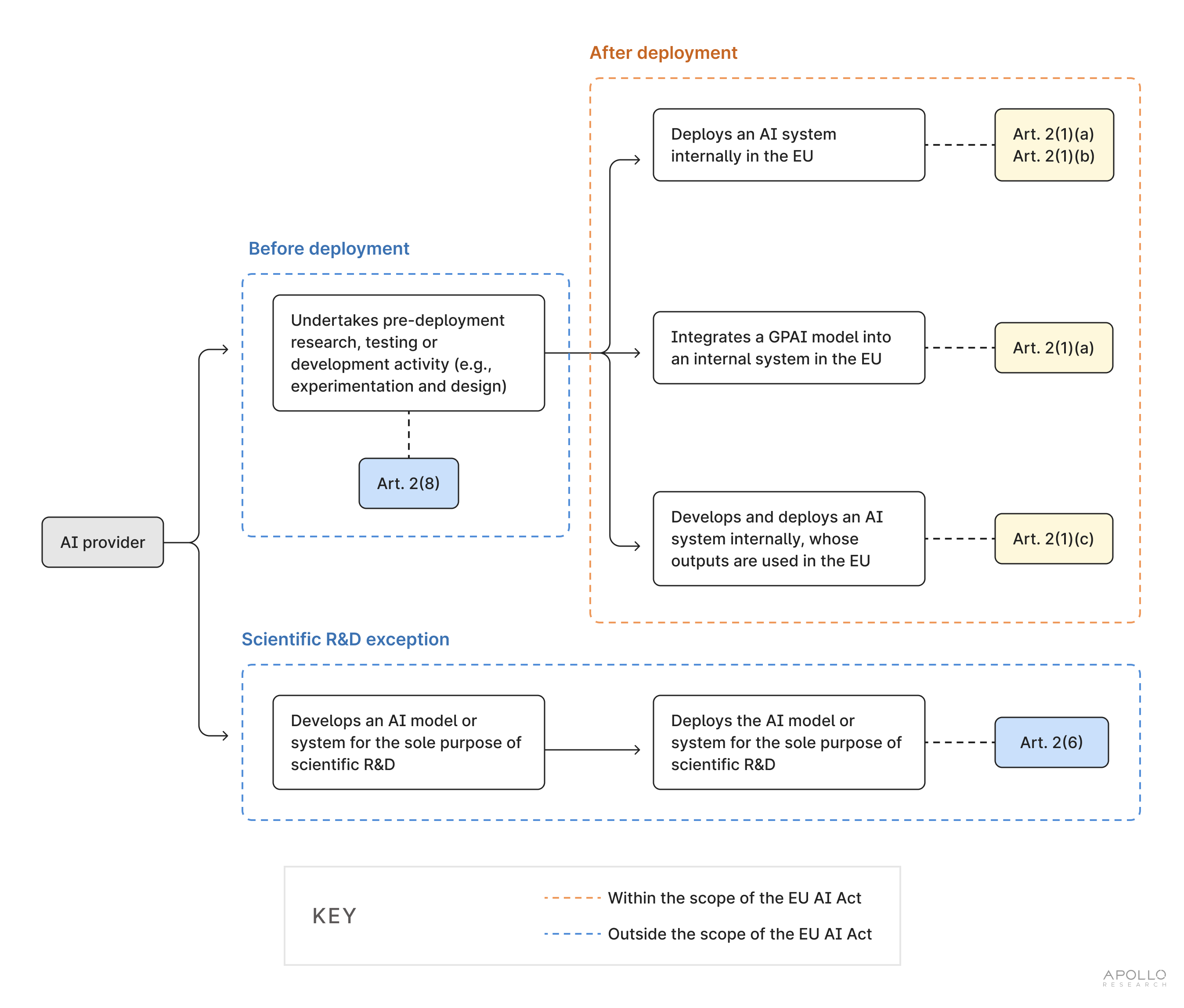}
\end{figure}

\textbf{Figure 1}: Visual representation of the scope of the AI Act in relation to AI models and systems deployed internally within AI providers, under a joint interpretation of Articles 2(1), 2(6), 2(8), 3(1), 3(3), 3(4), 3(9), 3(10), 3(11), 3(12), 3(63), and Recitals 12, 13, 21, 25, 97, 109, and 110 of the AI Act.

\begin{itemize}
  \item In yellow: internal AI models and systems arguably falling \emph{within} the scope of the AI Act under Article 2(1)(a)-(c) and Recital 97 (\emph{see} \S{}\S{}2.1.1-2.1.4 and \S{}2.2 above).
  \item In blue: the activities preceding internal deployment (\emph{see} \S{}3.2 above) and AI models or systems specifically developed and deployed for the sole purpose of scientific R\&D (\emph{see} \S{}3.1 above), arguably falling \emph{outside} the scope of the AI Act under Article 2(8) and Article 2(6) respectively.
\end{itemize}

In summary, \textbf{the deployment of an AI system, either \emph{within} or outside of an AI provider, triggers the application of the AI Act} (Figure 1, yellow box ``After deployment;'' \S{}\S{}2.1.1-2.1.4). This is clarified by \textbf{Article 2(1)(a)-(c)}, which sets the clear-cut rule that making available or using an AI system or its outputs in the EU triggers the application of the AI Act, and, in doing so, does not differentiate between internal systems and external systems.

\textbf{GPAI models trigger the application of the AI Act once they are integrated into an AI system (whether internal or external) that is put into service, or once they are independently placed on the market} (Figure 1, yellow box ``After deployment;'' \S{}2.2). This is clarified by \textbf{Article 2(1)(a)} and \textbf{Recital 97}, according to which a GPAI model counts as being placed on the market once it is integrated into an AI system. System integration is necessary or at least beneficial for most of the possible internal uses of a GPAI model, which in turn narrows down the number of cases in which an internal GPAI model is not covered by the AI Act.

\textbf{Internal AI models or systems specifically developed and deployed for the exclusive use of scientific research and development fall outside of the scope of the AI Act} (Figure 1, blue box ``Scientific R\&D exception;'' \S{}3.1). This exception is laid out in \textbf{Article 2(6)}, according to which the AI Act ``does not apply to AI systems or AI models, including their output, specifically developed and put into service for the sole purpose of scientific research and development.'' The scope of this exception is narrow. It arguably only applies to AI models and systems that providers have trained and deployed internally for the exclusive purpose of scientific research and development. While the AI Act does not define ``scientific research and development'' it is possible to infer from \textbf{Recital 25} and the REACH Regulation that scientific R\&D is different from ``product-oriented'' R\&D. By contrast to purely scientific R\&D, any AI model or system that engages in R\&D that is product-oriented does not fall within the Article 2(6) exception.

Finally, \textbf{research and development activities that providers undertake before ``putting into service'' or ``placing on the market'' an AI model or system} (for instance, experimentation and planning) \textbf{do not trigger \emph{per se} the application of the AI Act} (Figure 1, blue box ``Before deployment;'' \S{}3.2). This is clarified by \textbf{Article 2(8)}, according to which the AI Act ``does not apply to any research, testing or development activity regarding AI systems or AI models \emph{prior to} their being placed on the market or put into service.''

\nocite{*}
\bibliography{References}

@article{stix2025internal,
  author       = {Stix, Charlotte and Carlsmith, Joe and Schiavone, Stephen and Christiano, Paul and others},
  title        = {{AI} Behind Closed Doors: A Primer on the Governance of Internal Deployment},
  journal      = {arXiv preprint arXiv:2504.12170},
  year         = {2025},
  url          = {https://arxiv.org/abs/2504.12170}
}

@article{stix2025causal,
  author       = {Stix, Charlotte and Sharkey, Lee and others},
  title        = {A Causal Framework for {AI} Regulation and Auditing},
  journal      = {arXiv preprint arXiv:2511.15846},
  year         = {2025},
  url          = {https://arxiv.org/pdf/2511.15846}
}

@article{shevlane2024model,
  author       = {Shevlane, Toby and Farquhar, Sebastian and Garfinkel, Ben and Phuong, Mary and Whittlestone, Jess and Leung, Jade and Kokotajlo, Daniel and Marchal, Nahema and Anderljung, Markus and Kolt, Noam and Ho, Lewis and Siddarth, Divya and Avin, Shahar and Hawkins, Will and Kim, Been and Gabriel, Iason and Bolina, Vijay and Clark, Jack and Bengio, Yoshua and Christiano, Paul and Dafoe, Allan},
  title        = {Model Evaluation for Extreme Risks},
  journal      = {arXiv preprint arXiv:2305.15324},
  year         = {2024},
  url          = {https://arxiv.org/pdf/2305.15324}
}

@misc{hubinger2023model,
  author       = {Hubinger, Evan and Denison, Carson and Mu, Jesse and Lambert, Mike and Tong, Meg and MacDiarmid, Monte},
  title        = {Model Organisms of Misalignment: The Case for a New Pillar of Alignment Research},
  year         = {2023},
  howpublished = {AI Alignment Forum},
  url          = {https://www.alignmentforum.org/posts/ChDH335ckdvpxXaXX/model-organisms-of-misalignment-the-case-for-a-new-pillar-of-1}
}

@article{hubinger2024sleeper,
  author       = {Hubinger, Evan and Denison, Carson and Mu, Jesse and Lambert, Mike and Tong, Meg and MacDiarmid, Monte and Lanham, Tamera and Ziegler, Daniel M. and Maxwell, Tim and Cheng, Newton and Jermyn, Adam and Askell, Amanda and Radhakrishnan, Ansh and Anil, Cem and Duvenaud, David and Ganguli, Deep and Barez, Fazl and Clark, Jack and Ndousse, Kamal and Sachan, Kshitij and Sellitto, Michael and Sharma, Mrinank and DasSarma, Nova and Grosse, Roger and Kravec, Shauna and Bai, Yuntao and Witten, Zachary and Favaro, Marina and Brauner, Jan and Karnofsky, Holden and Christiano, Paul and Bowman, Samuel R. and Graham, Logan and Kaplan, Jared and Mindermann, S{\"o}ren and Greenblatt, Ryan and Shlegeris, Buck and Schiefer, Nicholas and Perez, Ethan},
  title        = {Sleeper Agents: Training Deceptive {LLM}s that Persist Through Safety Training},
  journal      = {arXiv preprint arXiv:2401.05566},
  year         = {2024},
  url          = {https://arxiv.org/pdf/2401.05566}
}

@article{betley2025emergent,
  author       = {Betley, Jan and Tan, Daniel and Warncke, Niels and Sztyber-Betley, Anna and Bao, Xuchan and Soto, Mart{\'\i}n and Labenz, Nathan and Evans, Owain},
  title        = {Emergent Misalignment: Narrow Finetuning Can Produce Broadly Misaligned {LLM}s},
  journal      = {arXiv preprint arXiv:2502.17424},
  year         = {2025},
  url          = {https://arxiv.org/pdf/2502.17424}
}

@article{turner2025model,
  author       = {Turner, Alex and others},
  title        = {Model Organisms for Emergent Misalignment},
  journal      = {arXiv preprint arXiv:2506.11613},
  year         = {2025},
  url          = {https://arxiv.org/pdf/2506.11613}
}

@article{wang2025persona,
  author       = {Wang, Miles and others},
  title        = {Persona Features Control Emergent Misalignment},
  journal      = {arXiv preprint arXiv:2506.19823},
  year         = {2025},
  url          = {https://www.arxiv.org/pdf/2506.19823}
}

@article{hua2025steering,
  author       = {Hua, Wenjia and others},
  title        = {Steering Evaluation-Aware Language Models to Act Like They Are Deployed},
  journal      = {arXiv preprint arXiv:2510.20487},
  year         = {2025},
  url          = {https://arxiv.org/pdf/2510.20487}
}

@article{marks2025auditing,
  author       = {Marks, Samuel and Treutlein, Johannes and Bricken, Trenton and Lindsey, Jack and Marcus, Jonathan and Mishra-Sharma, Siddharth and Ziegler, Daniel and Perez, Ethan and Sharma, Mrinank and Denison, Carson and Xu, Fabien Roger and others},
  title        = {Auditing Language Models for Hidden Objectives},
  journal      = {arXiv preprint arXiv:2503.10965},
  year         = {2025},
  url          = {https://arxiv.org/pdf/2503.10965}
}

@article{sharkey2024open,
  author       = {Sharkey, Lee and Chughtai, Bilal and Batson, Joshua and Lindsey, Jack and Wu, Jeff and Bushnaq, Lucius and Goldowsky-Dill, Nicholas and Heimersheim, Stefan and Ortega, Alejandro and Bloom, Joseph and Biderman, Stella and Conmy, Arthur},
  title        = {Open Problems in Mechanistic Interpretability},
  journal      = {arXiv preprint arXiv:2501.16496},
  year         = {2024},
  url          = {https://arxiv.org/pdf/2501.16496}
}

@techreport{sharkey2023causal,
  author       = {Sharkey, Lee and Ghuidhir, Cl{\'\i}odhna N{\'\i} and Braun, Dan and Scheurer, J{\'e}r{\'e}my and Balesni, Mikita and Bushnaq, Lucius and Stix, Charlotte and Hobbhahn, Marius},
  title        = {A Causal Framework for {AI} Regulation and Auditing},
  institution  = {Apollo Research},
  year         = {2023},
  url          = {https://www.apolloresearch.ai/u/2025/11/A-Causal-Framework-for-AI-Regulation-and-Auditing-.pdf}
}

@article{gao2025weight,
  author       = {Gao, Leo and others},
  title        = {Understanding Neural Networks through Sparse Circuits},
  journal      = {arXiv preprint arXiv:2511.13653},
  year         = {2025},
  url          = {https://arxiv.org/pdf/2511.13653}
}

@article{cundy2025honesty,
  author       = {Cundy, Chris and Gleave, Adam},
  title        = {Training a Reward Hacker Despite Perfect Labels},
  journal      = {arXiv preprint arXiv:2505.13787},
  year         = {2025},
  url          = {https://arxiv.org/abs/2505.13787}
}

@article{vaxenburg2025whole,
  author       = {Vaxenburg, Roman and Siwanowicz, Igor and Merel, Josh and Robie, Alice A. and Morrow, Carmen and Novati, Guido and Hasanbegovic, Zinovia and Hu, Shanqing and Banerjee, Suraj and Sanders, Tessa M. and Lauer, Jessica and others},
  title        = {Whole-body Simulation of Realistic Fruit Fly Locomotion with Deep Reinforcement Learning},
  journal      = {Nature},
  year         = {2025},
  url          = {https://www.nature.com/articles/s41586-025-09029-4}
}

@article{atlas2012observation,
  author       = {{ATLAS Collaboration}},
  title        = {Observation of a New Particle in the Search for the {S}tandard {M}odel {H}iggs Boson with the {ATLAS} Detector at the {LHC}},
  journal      = {Physics Letters B},
  volume       = {716},
  number       = {1},
  pages        = {1--29},
  year         = {2012},
  url          = {https://arxiv.org/pdf/1207.7214}
}

@article{jumper2021highly,
  author       = {Jumper, John and Evans, Richard and Pritzel, Alexander and Green, Tim and Figurnov, Michael and Ronneberger, Olaf and Tunyasuvunakool, Kathryn and Bates, Russ and {\v{Z}}{\'\i}dek, Augustin and Potapenko, Anna and others},
  title        = {Highly Accurate Protein Structure Prediction with {AlphaFold}},
  journal      = {Nature},
  volume       = {596},
  number       = {7873},
  pages        = {583--589},
  year         = {2021},
  url          = {https://www.nature.com/articles/s41586-021-03819-2}
}

@techreport{bubeck2025early,
  author       = {Bubeck, S{\'e}bastien and others},
  title        = {Early Science Acceleration Experiments with {GPT-5}},
  institution  = {OpenAI},
  year         = {2025},
  url          = {https://cdn.openai.com/pdf/4a25f921-e4e0-479a-9b38-5367b47e8fd0/early-science-acceleration-experiments-with-gpt-5.pdf}
}

@article{lu2024aiscientist,
  author       = {Lu, Chris and Lu, Cong and Lange, Robert Tjarko and Foerster, Jakob and Clune, Jeff and Ha, David},
  title        = {The {AI} Scientist: Towards Fully Automated Open-Ended Scientific Discovery},
  journal      = {arXiv preprint arXiv:2408.06292},
  year         = {2024},
  url          = {https://arxiv.org/pdf/2408.06292}
}

@article{kwa2025measuring,
  author       = {Kwa, Thomas and West, Ben and Becker, Joel and Deng, Amy and Garcia, Katharyn and Hasin, Max and Jawhar, Sami and Kinniment, Megan and Rush, Nate and Arx, Sydney Von and Bloom, Ryan and Broadley, Thomas and Du, Haoxing and Goodrich, Brian and Jurkovic, Nikola and Miles, Luke Harold and Nix, Sea and Lin, Tao and Parikh, Neev and Rein, David and Sato, Lucas Jun Koba and Wei, Hjalmar and Xu, Lawrence Chan and Murphy, Catherine and Hadshar, Ronak and Vivian, Tarun and others},
  title        = {Measuring {AI} Ability to Complete Long Tasks},
  journal      = {arXiv preprint arXiv:2503.14499},
  year         = {2025},
  url          = {https://arxiv.org/abs/2503.14499}
}

@techreport{wijk2024metr,
  author       = {Wijk, Hjalmar and Lin, Tao and Becker, Joel and Jawhar, Sami and Parikh, Neev and Broadley, Thomas and Chan, Lawrence and Chen, Michael and Clymer, Joshua and Dhyani, Jai and Ericheva, Elena and Garcia, Katharyn and Goodrich, Brian and Jurkovic, Nikola and Kinniment, Megan and Lajko, Aron and Nix, Seraphina and Sato, Lucas and Saunders, William and Taran, Maksym and West, Ben and Barnes, Elizabeth},
  title        = {{RE-Bench}: Evaluating Frontier {AI} {R\&D} Capabilities of Language Model Agents against Human Experts},
  institution  = {METR},
  year         = {2024},
  url          = {https://metr.org/AI_R_D_Evaluation_Report.pdf}
}

@misc{behrouz2025nested,
  author       = {Behrouz, Ali and others},
  title        = {Nested Learning: The Illusion of Deep Learning Architectures},
  year         = {2025},
  url          = {https://abehrouz.github.io/files/NL.pdf}
}

@article{hacker2023regulating,
  author       = {Hacker, Philipp and Engel, Andreas and Mauer, Marco},
  title        = {Regulating {ChatGPT} and Other Large Generative {AI} Models},
  journal      = {arXiv preprint arXiv:2302.02337},
  year         = {2023},
  url          = {https://arxiv.org/pdf/2302.02337}
}

@article{finck2025science,
  author       = {Finck, Michele},
  title        = {The Science Exception in the {EU} {AI} Act},
  journal      = {GRUR International},
  volume       = {74},
  number       = {10},
  pages        = {903--915},
  year         = {2025},
  url          = {https://academic.oup.com/grurint/article/74/10/903/8238058}
}

@incollection{vaneecke2024scope,
  author       = {Van Eecke, Patrick and Regenhardt, Janina},
  title        = {Scope of the {AI} Act},
  booktitle    = {The {EU} Artificial Intelligence ({AI}) Act: A Commentary},
  publisher    = {Wolters Kluwer},
  year         = {2024},
  url          = {https://law-store.wolterskluwer.com/s/product/the-eu-artificial-intelligence-ai-act-a-commentary/01tPg000007gkK9IAI}
}

@article{almada2024extraterritorial,
  author       = {Almada, Marco and Radu, Anca},
  title        = {The Brussels Side-Effect: How the {AI} Act Can Reduce the Global Reach of {EU} Policy},
  journal      = {German Law Journal},
  volume       = {25},
  number       = {4},
  pages        = {646--662},
  year         = {2024},
  url          = {https://doi.org/10.1017/glj.2023.108}
}

@article{czerniawski2024extraterritorial,
  author       = {Czerniawski, Mariusz},
  title        = {The Extraterritorial Scope of the {EU} {AI} Act},
  journal      = {Computer Law \& Security Review},
  year         = {2024}
}

@article{boine2025regulating,
  author       = {Boine, Claire and Rolnick, David},
  title        = {Regulating Foundation Models in the {AI} Act: Frontier {AI}, Open Source, and Compute Thresholds},
  journal      = {Minnesota Journal of Law, Science \& Technology},
  year         = {2025},
  url          = {https://scholarship.law.umn.edu/cgi/viewcontent.cgi?article=1576&context=mjlst}
}

@misc{lanamaki2024european,
  author       = {Lanam{\"a}ki, Arto and others},
  title        = {The {E}uropean Union's Regulatory Challenge: Conceptualizing Purpose in Artificial Intelligence},
  year         = {2024},
  url          = {https://www.researchgate.net/publication/380319968}
}

@misc{acharya2025managing,
  author       = {Acharya, Ashwin and Delaney, Oscar},
  title        = {Managing Risks from Internal {AI} Systems},
  year         = {2025},
  url          = {https://static1.squarespace.com/static/64edf8e7f2b10d716b5ba0e1/t/687e324254b8df665abc5664/1753100867033/Managing+Risks+from+Internal+AI+Systems.pdf}
}

@techreport{openai2023gpt4card,
  author       = {{OpenAI}},
  title        = {{GPT-4} System Card},
  institution  = {OpenAI},
  year         = {2023},
  url          = {https://cdn.openai.com/papers/gpt-4-system-card.pdf}
}

@techreport{openai2025gpt5card,
  author       = {{OpenAI}},
  title        = {{GPT-5} System Card},
  institution  = {OpenAI},
  year         = {2025},
  url          = {https://arxiv.org/pdf/2601.03267}
}

@techreport{openai2026gpt55card,
  author       = {{OpenAI}},
  title        = {{GPT-5.5} System Card},
  institution  = {OpenAI},
  year         = {2026},
  url          = {https://deploymentsafety.openai.com/gpt-5-5/gpt-5-5.pdf}
}

@techreport{anthropic2025opus45card,
  author       = {{Anthropic}},
  title        = {{Claude Opus 4.5} System Card},
  institution  = {Anthropic},
  year         = {2025},
  url          = {https://assets.anthropic.com/m/64823ba7485345a7/Claude-Opus-4-5-System-Card.pdf}
}

@techreport{anthropic2026opus46card,
  author       = {{Anthropic}},
  title        = {{Claude Opus 4.6} System Card},
  institution  = {Anthropic},
  year         = {2026},
  url          = {https://www-cdn.anthropic.com/0dd865075ad3132672ee0ab40b05a53f14cf5288.pdf}
}

@techreport{anthropic2026opus47card,
  author       = {{Anthropic}},
  title        = {{Claude Opus 4.7} System Card},
  institution  = {Anthropic},
  year         = {2026},
  url          = {https://cdn.sanity.io/files/4zrzovbb/website/037f06850df7fbe871e206dad004c3db5fd50340.pdf}
}

@techreport{anthropic2026mythoscard,
  author       = {{Anthropic}},
  title        = {{Claude Mythos} Preview System Card},
  institution  = {Anthropic},
  year         = {2026},
  url          = {https://www-cdn.anthropic.com/08ab9158070959f88f296514c21b7facce6f52bc.pdf}
}

@techreport{deepmind2025gemini3card,
  author       = {{Google DeepMind}},
  title        = {{Gemini 3 Pro} Model Card},
  institution  = {Google DeepMind},
  year         = {2025},
  url          = {https://storage.googleapis.com/deepmind-media/Model-Cards/Gemini-3-Pro-Model-Card.pdf}
}

@techreport{deepmind2025gemini3fsf,
  author       = {{Google DeepMind}},
  title        = {{Gemini 3 Pro} Frontier Safety Framework Report},
  institution  = {Google DeepMind},
  year         = {2025},
  url          = {https://storage.googleapis.com/deepmind-media/gemini/gemini_3_pro_fsf_report.pdf}
}

@techreport{openai2025preparedness,
  author       = {{OpenAI}},
  title        = {Preparedness Framework, Version 2},
  institution  = {OpenAI},
  year         = {2025},
  url          = {https://cdn.openai.com/pdf/18a02b5d-6b67-4cec-ab64-68cdfbddebcd/preparedness-framework-v2.pdf}
}

@techreport{anthropic2025rsp,
  author       = {{Anthropic}},
  title        = {Responsible Scaling Policy},
  institution  = {Anthropic},
  year         = {2025},
  url          = {https://www-cdn.anthropic.com/872c653b2d0501d6ab44cf87f43e1dc4853e4d37.pdf}
}

@techreport{anthropic2025modelreport,
  author       = {{Anthropic}},
  title        = {Transparency Model Report},
  institution  = {Anthropic},
  year         = {2025},
  url          = {https://www.anthropic.com/transparency/model-report}
}

@misc{anthropic2026saboteur,
  author       = {{Anthropic}},
  title        = {Auditing an {AI} Model for Overt Sabotage},
  year         = {2026},
  howpublished = {Anthropic Alignment},
  url          = {https://alignment.anthropic.com/2026/auditing-overt-saboteur/}
}

@misc{openai2025sparsecircuits,
  author       = {{OpenAI}},
  title        = {Understanding Neural Networks through Sparse Circuits},
  year         = {2025},
  howpublished = {OpenAI Blog},
  url          = {https://openai.com/index/understanding-neural-networks-through-sparse-circuits/}
}

@misc{openai2025progress,
  author       = {{OpenAI}},
  title        = {{AI} Progress and Recommendations},
  year         = {2025},
  howpublished = {OpenAI Blog},
  url          = {https://openai.com/index/ai-progress-and-recommendations/}
}

@techreport{deepmind2025fsf,
  author       = {{Google DeepMind}},
  title        = {Frontier Safety Framework, Version 3},
  institution  = {Google DeepMind},
  year         = {2025},
  url          = {https://storage.googleapis.com/deepmind-media/DeepMind.com/Blog/strengthening-our-frontier-safety-framework/frontier-safety-framework_3.pdf}
}

@misc{deepmind2025alphaevolve,
  author       = {{Google DeepMind}},
  title        = {{AlphaEvolve}: A {Gemini}-Powered Coding Agent for Designing Advanced Algorithms},
  year         = {2025},
  howpublished = {Google DeepMind Blog},
  url          = {https://deepmind.google/blog/alphaevolve-a-gemini-powered-coding-agent-for-designing-advanced-algorithms/}
}

@techreport{meta2026scaling,
  author       = {{Meta}},
  title        = {Advanced {AI} Scaling Framework, Version 2},
  institution  = {Meta},
  year         = {2026},
  url          = {https://ai.meta.com/static-resource/Meta_Advanced-AI-Scaling-Framework-v2}
}

@techreport{metr2025elements,
  author       = {{METR}},
  title        = {Common Elements of Frontier {AI} Safety Policies},
  institution  = {METR},
  year         = {2025},
  url          = {https://metr.org/common-elements.pdf}
}

@misc{metr2025dangerous,
  author       = {{METR}},
  title        = {{AI} Models Can Be Dangerous Before Public Deployment},
  year         = {2025},
  howpublished = {METR Blog},
  url          = {https://metr.org/blog/2025-01-17-ai-models-dangerous-before-public-deployment/}
}

@misc{metr2026measuring,
  author       = {{METR}},
  title        = {Measuring {AI} Ability to Complete Long Tasks},
  year         = {2026},
  howpublished = {METR Blog},
  url          = {https://metr.org/blog/2025-03-19-measuring-ai-ability-to-complete-long-tasks/}
}

@misc{amodei2024machines,
  author       = {Amodei, Dario},
  title        = {Machines of Loving Grace},
  year         = {2024},
  howpublished = {Personal essay},
  url          = {https://www.darioamodei.com/essay/machines-of-loving-grace}
}

@misc{adler2025unmonitored,
  author       = {Adler, Steven},
  title        = {{AI} Companies' Unmonitored Internal Use of {AI}},
  year         = {2025},
  howpublished = {Substack},
  url          = {https://stevenadler.substack.com/p/ai-companies-unmonitored-internal}
}

@misc{kokotajlo2025ai2027,
  author       = {Kokotajlo, Daniel and Alexander, Scott and Larsen, Thomas and Lifland, Eli and Dean, Romeo},
  title        = {{AI} 2027},
  year         = {2025},
  url          = {https://ai-2027.com/}
}

@misc{davidson2025incorporate,
  author       = {Davidson, Tom},
  title        = {How Can {AI} Labs Incorporate Risks from {AI} Accelerating {AI} Progress?},
  year         = {2025},
  howpublished = {Forethought},
  url          = {https://www.forethought.org/research/how-can-ai-labs-incorporate-risks-from-ai-accelerating-ai-progress-into}
}

@misc{davidson2025three,
  author       = {Davidson, Tom and Hadshar, Rose and MacAskill, Will},
  title        = {Three Types of Intelligence Explosion},
  year         = {2025},
  howpublished = {Forethought},
  url          = {https://www.forethought.org/research/three-types-of-intelligence-explosion}
}

@misc{eth2025software,
  author       = {Eth, Daniel and Davidson, Tom},
  title        = {Will {AI} {R\&D} Automation Cause a Software Intelligence Explosion?},
  year         = {2025},
  howpublished = {Forethought},
  url          = {https://www.forethought.org/research/will-ai-r-and-d-automation-cause-a-software-intelligence-explosion}
}

@misc{altman2025tweet,
  author       = {Altman, Sam},
  title        = {Post on automated {AI} researcher timelines},
  year         = {2025},
  howpublished = {X (Twitter)},
  url          = {https://x.com/sama/status/1983584366547829073}
}

@misc{guardian2025kaplan,
  author       = {{The Guardian}},
  title        = {Interview with {J}ared {K}aplan},
  year         = {2025},
  howpublished = {The Guardian},
  url          = {https://www.theguardian.com/technology/ng-interactive/2025/dec/02/jared-kaplan-artificial-intelligence-train-itself}
}

@misc{techcrunch2025altman,
  author       = {{TechCrunch}},
  title        = {{S}am {A}ltman Says {OpenAI} Will Have a Legitimate {AI} Researcher by 2028},
  year         = {2025},
  howpublished = {TechCrunch},
  url          = {https://techcrunch.com/2025/10/28/sam-altman-says-openai-will-have-a-legitimate-ai-researcher-by-2028/}
}

@techreport{intlsafetyreport2025,
  author       = {Bengio, Yoshua and others},
  title        = {International Scientific Report on the Safety of Advanced {AI}},
  institution  = {International {AI} Safety Report Secretariat},
  year         = {2025},
  url          = {https://internationalaisafetyreport.org/sites/default/files/2025-10/international_ai_safety_report_2025_english.pdf}
}

@techreport{echa2017ppord,
  author       = {{European Chemicals Agency}},
  title        = {Guidance on Scientific Research and Development ({SR\&D}) and Product and Process Orientated Research and Development ({PPORD})},
  institution  = {ECHA},
  year         = {2017},
  url          = {https://echa.europa.eu/documents/10162/2324906/ppord_en.pdf/22a12900-ad27-454c-aedd-82972ef2f675}
}

@techreport{eucommission2025aidef,
  author       = {{European Commission}},
  title        = {Commission Guidelines on the Definition of an {AI} System Established by Regulation ({EU}) 2024/1689 ({AI} Act)},
  institution  = {European Commission},
  year         = {2025},
  url          = {https://digital-strategy.ec.europa.eu/en/library/commission-publishes-guidelines-ai-system-definition-facilitate-first-ai-acts-rules-application}
}

@techreport{eucommission2025prohibited,
  author       = {{European Commission}},
  title        = {Commission Guidelines on Prohibited {AI} Practices Established by Regulation ({EU}) 2024/1689 ({AI} Act)},
  institution  = {European Commission},
  year         = {2025},
  url          = {https://digital-strategy.ec.europa.eu/en/library/commission-publishes-guidelines-prohibited-artificial-intelligence-ai-practices-defined-ai-act}
}

@techreport{eucommission2025gpai,
  author       = {{European Commission}},
  title        = {Guidelines on the Scope of the Obligations for General-Purpose {AI} Models Established by Regulation ({EU}) 2024/1689 ({AI} Act)},
  institution  = {European Commission},
  year         = {2025},
  url          = {https://digital-strategy.ec.europa.eu/en/library/guidelines-scope-obligations-providers-general-purpose-ai-models-under-ai-act}
}

@techreport{eucommission2022blueguide,
  author       = {{European Commission}},
  title        = {The `Blue Guide' on the Implementation of {EU} Product Rules 2022},
  institution  = {European Commission},
  year         = {2022},
  url          = {https://eur-lex.europa.eu/legal-content/EN/TXT/PDF/?uri=CELEX:52022XC0629(04)}
}

@techreport{eucommission2025cop,
  author       = {{European Commission}},
  title        = {Code of Practice for General-Purpose {AI} Models},
  institution  = {European Commission},
  year         = {2025},
  url          = {https://digital-strategy.ec.europa.eu/en/policies/contents-code-gpai}
}

@misc{aiact2024,
  author       = {{European Parliament and Council}},
  title        = {Regulation ({EU}) 2024/1689 of 13 {J}une 2024 Laying Down Harmonised Rules on Artificial Intelligence ({AI} Act)},
  year         = {2024},
  howpublished = {Official Journal of the European Union},
  url          = {https://eur-lex.europa.eu/legal-content/EN/TXT/PDF/?uri=OJ:L_202401689}
}

@misc{gdpr2016,
  author       = {{European Parliament and Council}},
  title        = {Regulation ({EU}) 2016/679 of 27 {A}pril 2016 ({G}eneral {D}ata {P}rotection {R}egulation)},
  year         = {2016},
  howpublished = {Official Journal of the European Union},
  url          = {https://eur-lex.europa.eu/eli/reg/2016/679/oj}
}

@misc{reach2006,
  author       = {{European Parliament and Council}},
  title        = {Regulation ({EC}) 1907/2006 of 18 {D}ecember 2006 Concerning the Registration, Evaluation, Authorisation and Restriction of Chemicals ({REACH})},
  year         = {2006},
  howpublished = {Official Journal of the European Union},
  url          = {https://eur-lex.europa.eu/legal-content/EN/TXT/PDF/?uri=CELEX:32006R1907}
}

@misc{tfeu2016,
  author       = {{European Union}},
  title        = {Consolidated Version of the Treaty on the Functioning of the {E}uropean {U}nion, Article 179},
  year         = {2016},
  url          = {https://eur-lex.europa.eu/legal-content/EN/TXT/HTML/?uri=CELEX:12016E179}
}

@misc{eo14363,
  author       = {Trump, Donald J.},
  title        = {Executive Order 14363 of {N}ovember 24, 2025: Launching the {G}enesis {M}ission},
  year         = {2025},
  howpublished = {The White House},
  url          = {https://www.whitehouse.gov/presidential-actions/2025/11/launching-the-genesis-mission/}
}

@misc{nsm2024ai,
  author       = {Biden, Joseph R.},
  title        = {Memorandum on Advancing the {U}nited {S}tates' Leadership in Artificial Intelligence; Harnessing Artificial Intelligence to Fulfill National Security Objectives; and Fostering the Safety, Security, and Trustworthiness of Artificial Intelligence},
  year         = {2024},
  howpublished = {The White House},
  url          = {https://bidenwhitehouse.archives.gov/briefing-room/presidential-actions/2024/10/24/memorandum-on-advancing-the-united-states-leadership-in-artificial-intelligence-harnessing-artificial-intelligence-to-fulfill-national-security-objectives-and-fostering-the-safety-security/}
}

@misc{casb53,
  author       = {{California State Legislature}},
  title        = {Senate Bill No.~53: {T}ransparency in Frontier Artificial Intelligence Act},
  year         = {2025},
  url          = {https://leginfo.legislature.ca.gov/faces/billTextClient.xhtml?bill_id=202520260SB53}
}

@misc{nobel2024chemistry,
  author       = {{The Nobel Prize}},
  title        = {The Nobel Prize in Chemistry 2024},
  year         = {2024},
  url          = {https://www.nobelprize.org/prizes/chemistry/2024/press-release/}
}

@misc{cern,
  author       = {{CERN}},
  title        = {The Large Hadron Collider},
  year         = {2024},
  note         = {Accessed via \url{https://home.web.cern.ch}},
  url          = {https://home.web.cern.ch/science/accelerators/large-hadron-collider}
}

@misc{hbp2023,
  author       = {{Human Brain Project}},
  title        = {The Human Brain Project Ends: What Has Been Achieved?},
  year         = {2023},
  url          = {https://www.humanbrainproject.eu/en/follow-hbp/news/2023/09/28/human-brain-project-ends-what-has-been-achieved/}
}

@misc{isomorphic2023,
  author       = {{Isomorphic Labs}},
  title        = {A Glimpse of the Next Generation of {AlphaFold}},
  year         = {2023},
  url          = {https://www.isomorphiclabs.com/articles/a-glimpse-of-the-next-generation-of-alphafold}
}
\addcontentsline{toc}{section}{References}

\end{document}